\documentclass[twocolumn,nofootinbib,amsmath,prd,aps,superscriptaddress,tightenlines,preprintnumbers]{revtex4}
\usepackage{graphicx}
\usepackage{epstopdf}
\usepackage{amsfonts}
\usepackage{bm}
\usepackage{epsfig}
\usepackage{graphics}
\usepackage{cancel}
\usepackage{xspace}

\usepackage[usenames,dvipsnames]{color}
     \definecolor{hgreen}{rgb}{0,.3,0}
     \definecolor{hred}{rgb}{.3,0,0}
     \definecolor{hblue}{rgb}{0,0,.3}
     \definecolor{LightGray}{gray}{0.95}
\usepackage[
            colorlinks=true,
            linkcolor=hblue,
            citecolor=hgreen,
            filecolor=hblue,
            urlcolor=hred]{hyperref}

\def\be{\begin{equation}}
\def\ee{\end{equation}}
\def\beq{\begin{equation}}
\def\eeq{\end{equation}}
\newcommand{\lsim}{\mathrel{\rlap{\lower4pt\hbox{\hskip1pt$\sim$}}
    \raise1pt\hbox{$<$}}}         

\newcommand{\pb}{{\rm\,pb}}
\def\AFB{$A_{FB}^{t \bar t}$}
\def\AC{A_{\rm C}}

\def\q{{\cal Q}}

\begin{document}

\title{
Stealth QCD-like Strong Interactions and the $t\bar t$ Asymmetry
}

\def\Cincy{Department of Physics, University of Cincinnati, Cincinnati, Ohio 45221,USA}

\author{Joachim Brod}
\email[]{joachim.brod@uc.edu}
\affiliation{\Cincy}

\author{Jure Drobnak}
\email[]{jure.drobnak@tum.de}
\affiliation{Technische Universit\"at M\"unchen, TUM Institute for Advanced Study, Lichtenbergstrasse 2a, D-85748 Garching, Germany}

\author{Alexander L. Kagan}
\email[]{kaganalexander@gmail.com} 
\affiliation{\Cincy}

\author{Emmanuel Stamou}
\email[]{emmanuel.stamou@weizmann.ac.il}
\affiliation{Department of Particle Physics \& Astrophysics, Weizmann Institute of Science, Rehovot 76100, Israel}
\affiliation{Technische Universit\"at M\"unchen, Excellence Cluster Universe, Boltzmannstrasse 2, D-85748 Garching, Germany}

\author{Jure Zupan}
\email[]{zupanje@ucmail.uc.edu}
\affiliation{\Cincy}

\begin{abstract}
We show that a new strongly interacting sector can produce large
enhancements of the $t\bar t$ asymmetries at the Tevatron. The
Standard Model is extended by a new vector-like flavor triplet of
fermions and one heavy scalar, all charged under a hypercolor gauge
group ${\rm SU}(3)_{\rm HC}$. This simple extension results in a
number of new resonances. The predictions of our model are rather
rigid once a small number of UV parameters are fixed, since all the
strong dynamics can be directly taken over from our understanding of
QCD dynamics. Despite the rather low hypercolor confinement scale of
$\sim100$\,GeV, the new strongly interacting sector is stealth. It is
shielded from present direct and indirect NP searches since the light
resonances are QCD singlets, whereas the production of the heavier QCD
colored resonances leads predominantly to high multiplicity final
states. Improved searches can potentially be devised using top tagged
final states or decays into a small number of hypercolor pions.
\end{abstract}

\maketitle
\newpage
\pdfbookmark[1]{Introduction}{introduction}

{\bf Introduction.} Strongly interacting theories with a low
confinement scale, e.g., $f_\pi \lsim 100$\,GeV, are a particularly
interesting possibility for new physics in the LHC era, due to the
large number of resonances that could be experimentally accessible.
In view of the recent discovery of the Higgs-like particle an
interesting framework in which a low confinement scale is motivated by
electroweak symmetry breaking is ``bosonic technicolor", where the
vacuum expectation value of a fundamental Higgs is induced from
technicolor (TC) dynamics~\cite{Simmons:1988fu, Samuel:1990dq,
  Dine:1990jd, Carone:1992rh, Carone:1993xc, kagantalk, Antola:2009wq,
  Azatov:2011ht, Azatov:2011ps}, and supersymmetry is introduced to
protect the Higgs mass \cite{Samuel:1990dq, Dine:1990jd, kagantalk,
  Azatov:2011ht, Azatov:2011ps}.

More generally, several classes of strongly interacting theories with
a low confinement scale and distinct phenomenologies have been
proposed, which are not directly linked to electroweak symmetry
breaking.  In ``hidden valley" models
\cite{Strassler:2006im,Han:2007ae}, the new strongly interacting
fermions which undergo confinement are neutral with respect to the
Standard Model (SM) interactions, thus effectively hiding their bound
states at colliders.  In ``quirk" models, the confining fermions or
quirks are taken to have color or electroweak charges, and have masses
that are much heavier than the new strong interaction scale
\cite{Okun:1980kw, Okun:1980mu, Kang:2008ea, Kribs:2009fy,
  Burdman:2008ek, Harnik:2008ax, Martin:2010kk}.  They therefore form
long stable strings at colliders with exotic signatures that depend on
the quirk mass.  Finally, in Refs.~\cite{Kilic:2008pm, Kilic:2008ub,
  Kilic:2009mi, Kilic:2010et} collider signatures of ``vectorlike
confinement" models were studied, in which the confining ``hypercolor
quark" masses are small compared to the confinement scale, as in QCD.
Variants containing weak scale colored mesons already tend to be ruled
out by LHC and Tevatron data.

In this paper we show that potentially enhanced top quark
forward-backward asymmetries (\AFB) at the Tevatron could be a
manifestation of a new strong interaction with a low confinement scale
that is not directly related to electroweak symmetry breaking.  Among
the many proposals leading to large asymmetries~\cite{Kamenik:2011wt},
only a small subset satisfies all flavor constraints without
fine-tuning. The t-channel models have new vector or scalar resonances
with masses of ${\mathcal O}(200~{\rm GeV})$, transforming
non-trivially under flavor symmetries \cite{Grinstein:2011yv,
  Grinstein:2011dz, Gresham:2011fx}.

There are two possibilities for flavorful vector fields in a
renormalizable theory: either they are fundamental gauge bosons of a
flavor symmetry or they are composite. While theories with light
gauged flavor bosons are a logical possibility, flavor constraints
could be particularly challenging to satisfy, and they could require a
complicated and potentially fine-tuned Higgs sector, see e.g.\,
Refs.~\cite{Grinstein:2010ve,Ko:2011vd}.
We thus explore the second option, which implies a strong
interaction confinement scale of $\sim 100$\,GeV.  Surprisingly, this
possibility is not excluded by existing collider searches.

We build an explicit model with a confining hypercolor (HC) gauge
group, ${\rm SU}(3)_{\rm HC}$.  The hypercolored matter consists of three
``light" flavors of vectorlike fermion ``HC quarks" that are neutral
with respect to the SM gauge interactions (they will be identified
with the flavors of the ordinary right-handed up quarks), with masses
of $\sim 3-30$\,GeV lying well below the strong interaction scale, like
the $u,d,s$ quarks in QCD.  A hallmark of this model is the existence,
following confinement, of new SM singlet resonances with masses
between $60$ and $300$\,GeV organized in flavor nonets of pseudoscalar,
vector, axial-vector and higher mass resonances. They are the
equivalent of the QCD resonances but with HC scale confinement
dynamics. There is also a ``heavy" flavor singlet fundamental ``HC
scalar", ${\cal S}$, with a mass of $\sim 500$\,GeV lying well above
the strong interaction scale.  It is an electrically charged QCD
triplet which decays before it can hadronize.

Remarkably, while the new states are relatively light and can lead to
large changes in the Tevatron $t\bar t$ asymmetries, the current LHC
and Tevatron searches are not yet sensitive to their production. The
reason is that in our model the light HC resonances are color singlets
and thus have relatively small cross sections. The only QCD colored
new states are the heavier elementary HC scalars that carry both QCD
and HC charge.  However, their detection is challenging, as their
production results in high multiplicity events, for instance $pp\to n
\pi_{\rm HC}+2j$ with $n$ large, and with the HC pions, $\pi_{\rm
  HC}$, decaying to two jets.

In the limit of a large mass for the HC scalar ${\cal S}$ our model
can effectively be thought of as a confining hidden valley model
\cite{Strassler:2006im, Han:2007ae}. In both cases the composite
resonances are not charged under the SM. However, in our case the
interaction with the SM is not through a $Z'$ as in
Refs.~\cite{Strassler:2006im, Han:2007ae}, but through Yukawa-like
interactions involving a HC quark, the HC scalar ${\cal S}$ and an
ordinary right-handed up quark. Unlike in the case of the hidden
valley, the HC pions decay to quark pairs, and we thus have no or very
little missing $E_T$ and/or leptons in the events.  The second
difference between our model and hidden valley models is that the
couplings of the new states to the SM are large. Thus, virtual
exchanges of these states can lead to observable consequences, such as
significant changes to the $t\bar t$ asymmetry at the Tevatron.

\begin{figure}
  \centering
  \includegraphics[width=0.4\columnwidth]{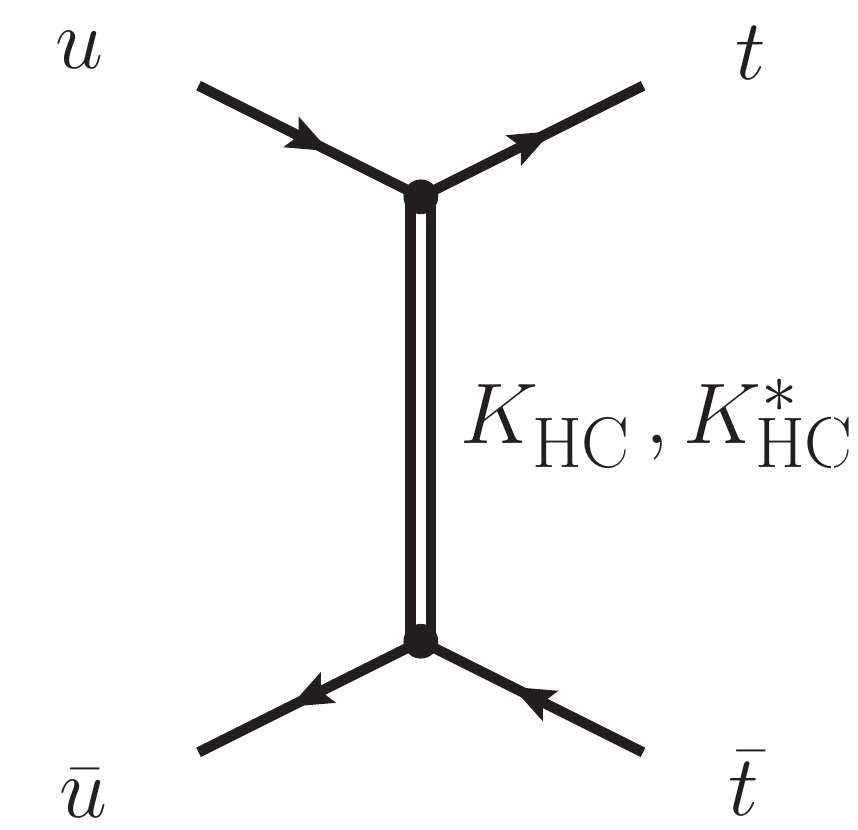}  
  \caption{Feynman diagram generating a contribution to the $t\bar t$
    asymmetry via a t-channel exchange of $K$ and $K^*$ resonances.
  \label{fig:tchannelK}}
\end{figure}

The two most important NP contributions to the $t\bar t$ asymmetry are
the exchanges of the HC vector resonance $K^*_{\rm HC}$ and the
pseudoscalar resonance $K_{\rm HC}$, by virtue of their flavor
off-diagonal couplings to the $u$ and $t$ quarks, see
Figure~\ref{fig:tchannelK}.
Our model is thus a realization of the effective t-channel models that
have been discussed in the literature, but with both vector and
(pseudo)scalar exchanges, rather than only one of these.  The $K^*$
acts as the $Z'$ in effective t-channel models~\cite{Jung:2009jz} and
the $K$ as the (pseudo)scalar state~\cite{Cui:2011xy, Blum:2011fa}.
At the LHC, the associated production of $K_{\rm HC}^*$ and $K_{\rm
  HC}$ in the $t\bar t j$ final state is also important for reducing
the charge asymmetry $\AC$, as stressed for $Z'$ models
in~\cite{Drobnak:2012rb}.  In order to avoid other constraints,
e.g. bounds on top-jet resonance production at the LHC, the branching
ratio of the decay $K_{\rm HC}^* \to q \bar t$ should be of order
$25\%$. This is achieved naturally in our model since the $K_{\rm
  HC}^*$ decays dominantly into pairs of pseudoscalar mesons, in
analogy with the decay $K^* \to K\pi$ in QCD.

Note that in order to have large $t\bar t$ asymmetry a relatively
large Yukawa coupling between the HC and the SM fields is needed.  For
the considered values of the Yukawa couplings the two-loop
renormalization-group equations (RGEs) suggest the existence of a
strongly interacting UV fixed point.  Under this assumption our theory
can thus be extended to arbitrarily high scales in the UV.

The paper is organized as follows. In Section~\ref{section:Set-up} we
introduce our simple extension of the Standard Model, in
Section~\ref{sec:resonances} we discuss in detail the mass spectra and
interactions of the resulting resonances, while
Section~\ref{sec:benchmarks} covers the LHC and Tevatron
phenomenology, including the $t\bar t$
asymmetries. Section~\ref{sec:EWK} covers precision electroweak
constraints, and perturbations of the Higgs properties.  Future
searches are covered in Section~\ref{sec:signals}, including a brief
discussion of the lightest HC baryon and dark matter direct detection
experiments.  Our conclusions are collected in
Section~\ref{sec:Conclusions}.  Three appendices contain further
details on translating information from QCD to the parameters of the
effective HC interaction Lagrangians.

\section{The set-up.\label{section:Set-up}}%
QCD provides the prototype for a confining theory with a spectrum that
contains flavorful vector mesons.  Using QCD as a guide, we introduce
an asymptotically free confining ${\rm SU}(N)_{\rm HC}$ hypercolor (HC)
gauge group. The anomaly-free matter content consists of the SM and
three copies of vectorlike hypercolor quarks $\q_{L i}
,\q_{R i} $ ($i=1,2,3$) and a flavor-singlet hypercolor scalar ${\cal
  S}$, transforming as 
\begin{equation} 
\begin{split}
  \q_{L i }&\sim (N,1,1,a)\,,\\
  \q_{R i} &\sim(N,1,1,a)\,,\\
  {\cal S} &\sim (\overline{N},3,1,b)\,,
 \label{hypercolormatter}
\end{split}
\end{equation}
with respect to the gauge symmetry ${\rm SU}(N)_{\rm HC} \times {\rm SU}(3)_{\rm C}
\times {\rm SU}(2)_L \times {\rm U}(1)_Y $.  The hypercharge 
assignments satisfy $a+b = 2/3$.  The choice $a=0, b=2/3$ is
phenomenologically favored, as we will see below, and will be used 
in the main part of the paper.
 
The most general renormalizable New Physics (NP) Lagrangian is given by
\begin{equation}\label{eq:LagHC}
\begin{split}
  -{\cal L}_{\rm NP}=& \big({\bf h}_{ij} \,\bar u_{Ri}\,\q_{Lj}
  \,{\cal S}\,  + \,{\rm h.c.}\big)\, + \,{\bf m_\q }_{ij}
  \,\bar{\q}_i \,\q_j\\ 
  & \,+ \, \mu_{\cal S}^2 \,{\rm Tr}({\cal S}^\dagger {\cal S})\, + \lambda_1 {\rm Tr}({\cal S}^\dagger {\cal S})^2 \\
  & + \lambda_2  {\rm Tr}({\cal S}^\dagger {\cal S} {\cal S}^\dagger
        {\cal S})+ \lambda_H H^\dagger H \,{\rm Tr}({\cal S}^\dagger
        {\cal S})\, ,
\end{split}
\end{equation}
plus the kinetic energy terms, where $H$ is the SM Higgs and $u_{Ri}$ 
the SM right-handed (RH) up quarks. 
The quartic interactions are not relevant for this work.  However, we
do require that the couplings $\lambda_{1,2}, \lambda_H$ do not lead
to a non-vanishing vacuum expectation value (vev) for ${\cal S}$.  The
total mass of the scalar ${\cal S}$ is $m_{\cal S}^2 = \mu_{\cal S}^2
+ \lambda_H v^2 /2$, where $v / \sqrt{2}$ is the SM Higgs vev.  In the
context of naturalness, we can imagine that the scalar ${\cal S}$ is
actually composite, or we can invoke supersymmetry above $O( 1~{\rm TeV})$ to
protect its mass.
 
In the absence of Yukawa interactions and HC quark masses the theory 
respects the global symmetry group ${\rm G_F}= {\rm U}(3)_{U_R} \times
{\rm U}(3)_{D_R} \times {\rm U}(3)_{Q_L}$, like the SM.
Under the ${\rm U}(3)_{U_R}$ symmetry both $({u_R}_ 1 , {u_R}_2 , {u_R}_3)$ and
$(\q_1, \q_2 , \q_3) $ transform as flavor triplets.
In the SM the main sources of flavour breaking are the top and bottom 
Yukawa couplings that break $\rm G_F$ to its subgroup 
\beq {\rm H_F}=
{\rm U}(2)_{U_R}\times {\rm U}(2)_{D_R}\times {\rm U}(2)_{Q_L}\times {\rm U}(1)_3\,.\eeq
Here we assume that the NP interactions also respect ${\rm H_F}$, i.e. 
the new Yukawa couplings
and mass terms are of the form
\begin{equation}
\label{eq:handm} 
  {\bf h } = {\rm diag } (h_1 , h_1 , h_3 ),~{\bf m_\q } = {\rm diag}
  ({m_{\q}}_1 , {m_{\q}}_1 , {m_{\q}}_3 ).  
\end{equation}

The breaking of $\rm H_F$ in the SM is due to the light-quark Yukawa
couplings and the CKM mixing angles, and is thus small. We will assume
that this breaking continues to be small in our model.
The approximate ${\rm U}(2)_{U_R}$ symmetry protects against dangerous 
flavor violation. We assume that its breaking is small, e.g., of MFV type.  Thus, new
contributions to $D_0$--$\overline{D_0}$ mixing as well as single and same-sign
top production 
are negligible.\footnote{Alternatively, one could entertain the possibility 
of an abelian ${\rm U}(1)_{\rm H}$ horizontal symmetry, with diagonal but fully non-degenerate 
${\bf h}$ and ${\bf m_\q}$ entries.}

We stress the following: {\it (i)} the fact that the hypercolor matter
only couples to the RH up quarks is due to the choice of
representations.
For example, had we chosen a hypercharge assignment such that $a+b=-1/3$ 
in Eq.~\eqref{hypercolormatter}, the hypercolor quarks $\q_i $ would only 
couple to the RH SM down quarks.
Alternatively, they could couple to the left-handed (LH) quarks if the
$\q_i$'s were ${\rm SU}(2)_L$ doublets.
{\it (ii)} Our set-up is potentially compatible with generation of 
the quark mass and mixing hierarchies via spontaneous breaking of a 
horizontal non-abelian symmetry in the UV, e.g., ${\rm SU}(3)$, ${\rm SU}(2)$, 
or discrete non-abelian groups.\footnote{Again, abelian ${\rm U}(1)$ symmetries 
provide a possible alternative.}
In such a scenario the 
${\rm U}(2)_{U_R}$ global symmetry of the
hypercolor sector would be a consequence 
of the underlying
horizontal symmetry, under which {\it all of the quarks} transform.  
{\it (iii)} The flavor structure of the
resonance mass spectrum, to be discussed below, could provide a hint
for the existence of such a fundamental horizontal symmetry in the UV
(see also Ref.~\cite{Grossman:2007bd}).

\begin{table}
\begin{ruledtabular}
\begin{tabular}{ccc}
  HC Resonance & Mass & Decay Width\\\hline
  $\pi_{\rm HC}$ & 62 GeV & $4.0 \cdot 10^{-7}  \,m_\pi$ \\
  $K_{\rm HC}$ &  143 GeV & $5.5 \cdot 10^{-7}  \,m_K$ \\
  $\eta_{\rm HC}$ &  161 GeV & $1.3 \cdot 10^{-7}  \,m_\eta$ \\
  \hline
  $\rho_{\rm HC}$ &  177 GeV & $0.059 \,m_\rho$  \\
  $K^*_{\rm HC}$ &  211 GeV & $0.002  \,m_{K^*}$  \\
  $V_H[\phi_{\rm HC}]$ &  242 GeV & $8.0 \cdot 10^{-7} \,m_{V_H}$  \\
  $V_L[\omega_{\rm HC}]$ &  180 GeV & $0.001 \,m_{V_L}$  \\
  \hline
  $a_1^{\rm HC}$ & 273 GeV &  $0.23  \,m_{a_1}$\\
  $K_1^{\rm HC}$ & 295 GeV &  $0.057  \,m_{K_1}$\\
  $f_1^{\rm HC}$ & 280 GeV &  $0.002 \,m_{f_1}$\\
  $f_1'{}^{\rm HC}$ & 320 GeV &  $3.2 \cdot 10^{-5} \,m_{f_1'}$\\
\end{tabular}
\end{ruledtabular}
\caption{The spectrum of the HC resonances for our benchmark.
}
\label{tab:PG}
\end{table}

\section{Hypercolor resonances and interactions\label{sec:resonances}}%
In order to make use of the available information on nonperturbative
QCD dynamics, we take the HC gauge group to be ${\rm SU}(3)_{\rm HC}$.  The
HC condensates, resonance masses, and couplings are estimated via
naive dimensional analysis (NDA), vector meson dominance (VMD), and/or
scaling from QCD.
The $\q$ and ${\cal S}$ masses are taken to satisfy $m_{\q_i} \ll
\Lambda_{\chi}$ and $m_{\cal S} \gtrsim \Lambda_{\chi}$, where
$\Lambda_{\chi}\sim 4\pi f_\pi$ is the HC chiral symmetry breaking
scale (the motivation for this choice is given below, in
Section~\ref{subsec:compquarks}). We also introduce the scale
$\Lambda_{\rm HC}\sim {\mathcal O}({\rm few}) f_\pi$, which is the
equivalent of $\Lambda_{\rm QCD}$ in QCD. The phenomenology that we
are interested in is dominated by the lowest lying HC resonances. We
thus keep the following resonances in the description
\begin{itemize}
\item
the flavor octet of pseudo-Goldstone pseudoscalar resonances, $\pi^a_{\rm HC}$,
\item
the set of lowest lying vector, $\rho^a_{\rm HC}$,  and axial vector,
$a_{1,{\rm HC}}^a$, flavor nonet resonances.
\end{itemize}

In principle one could include additional HC resonances, e.g., the
${}^{1}P_1$ axial-vector multiplet (in QCD it contains the $b_1(1235)$
and $K_{1B}$). For simplicity, however, we ignore them in this
work. Our notation is directly borrowed from QCD: $\pi_{\rm HC}$ is
thus the equivalent of $\pi$ in QCD, etc. To shorten the notation we
will often drop the HC subscript.  For this reason in this paper the
QCD states always carry a QCD subscript or superscript.

To illustrate the effect on $t\bar t$ production and other collider
and low-energy observables we consider a benchmark in the parameter
space. The benchmark resonance masses and decay widths are given in Table~\ref{tab:PG}.
The underlying UV parameters, as well as resonance couplings and decay constants are 
listed in Table~\ref{tab:bmuv}.
The determination of the resonance properties
is described in detail below.

\subsection{Pseudo-Nambu-Goldstone bosons\label{subsec:pseudoGB}}%
In the ${\rm U}(3)_{U_R}$ symmetric limit, the $\q_i $ form equal
condensates, $\langle \bar \q \q \rangle \ne 0$, which break the HC
sector chiral symmetry to the diagonal subgroup, ${\rm SU}(3)_L \times
{\rm SU}(3)_R \to {\rm SU}(3)_V $. This gives rise to a flavor octet of
pseudo-Nambu-Goldstone bosons (pNGBs) $\pi^a_{\,\rm HC} $
($a=1,\dots,8$).  In NDA, $\Lambda_{\chi } \sim 4 \pi
f_{\pi}^{\,\rm HC}$ and
\begin{equation}
\label{eq:mpi} 
  \langle \bar \q \q \rangle \sim 4 \pi (f_{\pi}^{\,\rm HC})^3,\,~~~~ 
m^2_{\pi^a} \sim 8 \pi f_{\pi}^{\,\rm HC} \,m_\q \,,  
\end{equation}
where $f_\pi^{\,\rm HC}$ is the HC-pion decay constant, 
\begin{equation} 
  \langle \pi^a | \bar\q T^a \gamma_\mu \gamma_5 \q | 0 \rangle = - i
  f^{\rm HC}_\pi p_\mu \,. 
\end{equation}
The flavor octet Gell-Mann matrices are normalized as ${\rm Tr}[T^a
  T^b] = \delta^{ab}/2 $.  For $ \langle \bar \q \q \rangle$ we use
the recent lattice determination of the QCD condensate
\cite{McNeile:2012xh}, which gives $29.8 (f_\pi^{\rm QCD})^3$, instead
of the NDA estimate $4\pi (f_\pi^{\rm QCD})^3$ in Eq.~\eqref{eq:mpi}
(in our convention $f_\pi^{\rm QCD}\simeq 92$\,MeV).  Similarly, for
the pion mass we take $2\times 29.8$ instead of the factor $8\pi$ in
Eq.~\eqref{eq:mpi}. Requiring the vector resonances to have masses in
the phenomenologically favored range of approximately $200$\,GeV fixes
$f_\pi^{\rm HC}\!\simeq\! 20$\,GeV, cf. Eq.~\eqref{eq:scaling} below.
Thus, for $m_\q\! \sim\! {\mathcal O}(10)$\,GeV the masses of the
pseudoscalars are $m_\pi^{\rm HC}\!\sim\! {\mathcal O}(100)$\,GeV.

In our benchmark we take ${m_{\q}}_3\!\sim\! f_\pi$ and
${m_{\q}}_1\!\sim\! {m_{\q}}_3/10$.  The flavor symmetry breaking,
$m_{\q_1}\!\ne\! m_{\q_3}$ leads to mass splitting between the
$\pi\!\sim\! [\bar\q_{1,2}\q_{1,2}]$, $K\!\sim\! [\bar \q_{1,2}
  \q_3]$, and $\eta\!\sim\! [ \bar \q_1 \q_1+\bar \q_2 \q_2- 2\bar
  \q_3 \q_3]$, where the valence-quark content is given in square
brackets.  Details of the evaluation of their masses are given in
Appendix~\ref{app:fit:to:spectra}.  The pseudoscalar mass spectrum
also contains a heavier $\eta^\prime_{\rm HC}$ flavor singlet.  The
mass of the $\eta'_{\rm HC}$ is $m_{\eta'}\sim \Lambda_{\rm \chi}$,
i.e., $m_{\eta'}\!\sim {\mathcal O}(250)$\,GeV.  The $\eta_{\rm HC}'$
has flavor diagonal couplings to the SM quarks.  For the light quarks
these are suppressed by the light quark masses.  The $\eta_{\rm HC}'$
thus has a negligible impact on $t\bar t$ and dijet phenomenology, as
well as on the vector decay widths (due to its large mass). As such it
can be omitted from our discussion.  For simplicity we also neglect
``$\,\eta - \eta^\prime\,$'' mixing.

\begin{table}
\centering
\begin{tabular}{cccccc}
\hline\hline
Param.	& Value~~&~~Param.	& Value~~&~~Param. & Value\\
\hline
$M_{\chi}$[GeV]	& $171$ & $g_\rho$ & $4.88$ & $f_\pi$ & $20.4$\,GeV\\
$m_{\q_1}$[GeV]        		& $3.1$ & $g_{\rho\pi\pi}$ & $4.88$ & $f_\rho$ & $32.6$\,GeV  \\
$m_{\q_3}$[GeV]  & $30.5$& $g_{a_1}$ &$6.73$& $f_{a_1}$ & $37.1$\,GeV \\
$m_{\cal S}  $[GeV]   & $520$ & $g_{V_o}$ & 12.5& $f_{u',c'}$ & $53.5$\,GeV \\ 
$h_1     $      & $2.0$   &  $g_{V_s}$ & $5.10$& $f_{t'}$ & $52.4$\,GeV \\
$h_3     $      & $4.2$ & $g_A$ & $1.26$& & \\
\hline\hline
\end{tabular}
\caption{The UV parameters for the benchmark point (left two columns),
  the resulting HC resonance couplings (middle two columns), and the
  HC decay constants (last two columns). 
  }
\label{tab:bmuv}
\end{table}

\subsection{Vectors and axial vectors\label{subsec:vec:axial}}%
As in QCD, the $[\bar \q \q]$ bound states give rise to
vector and axial-vector resonances that are flavor nonets. 
We denote the lowest lying states by $\rho_{\,\rm HC}^a $ and $a_{1\,\rm HC}^a $
($a=1,\ldots,9$), respectively.  Here, $\rho^9_{\rm HC} $ and $a_{1\,\rm HC}^9$ 
are the flavor-singlet states.  Sometimes we will also use the generic notation of 
$V$ and $A$ for vector and axial-vector mesons.
 
We first discuss the properties of these resonances in the flavor-symmetric limit. 
The masses and decay constants  can be estimated via the approximate
scaling relations
\begin{equation}
\label{eq:scaling}
  {f^{\,\rm HC}_{\pi} \over f^{\,\rm QCD}_\pi } \sim {f^{\, \rm HC}_{\rho \,(a_1)} \over  f^{\,\rm QCD}_{\rho \,(a_1)} } \sim {m_{\rho \,(a_1)}^{\,{HC}} \over m^{\,\rm QCD}_{\rho \,(a_1)}}\, ,
\end{equation}
with $f_\pi^{\,\rm QCD} \simeq 92$\,MeV.
The $\rho_{\,\rm HC}^a $ and $a_{1\,\rm HC}^a $ decay constants are defined as 
\begin{align} 
  \langle\rho^a_{\rm HC}|\bar\q\,T^a\,\gamma_\mu\,\q|0\rangle &= -i  f^{\rm HC}_{\rho} m^{\rm HC}_{\rho}\epsilon_\mu\,,\label{rho:decay}\\
  \langle a_{1\,\rm HC}^a|\bar\q\,T^a\,\gamma_\mu\gamma_5\,\q|0\rangle &= -i f^{\rm HC}_{a_1}m^{\rm HC}_{a_1}\epsilon_\mu\,,
\end{align}
where the flavor-singlet matrix $T^9 \equiv I_{3\times 3}
/\sqrt{6}$. In QCD, the $\rho$ decay constant is $f^{\rm QCD} _\rho
\simeq 148 $\,MeV \cite{Beneke:2006hg}.  The $a_1$ decay constant
$f_{a_1}^{\rm QCD}$ is not well known.  For example, a light-cone
sum-rule determination yields $f_{a_1}^{\rm QCD} \simeq 168$\,MeV
\cite{Yang:2007zt}. Isgur et al.~\cite{Isgur:1988vm} made a
phenomenological determination from the $\tau^+ \to \nu_\tau \,\pi^+
\pi^+ \pi^- $ branching ratio.  Rescaling their result to the current
PDG branching ratio \cite{Beringer:1900zz} would yield $f_{a_1} \simeq
152$\,MeV.  (Note that the quoted values of the $\rho$ and $a_1$ decay
constants have been reduced by a factor $1/\sqrt{2}$ to conform to our
normalization for $f_\pi$.)  As an example, taking the sum-rule value
for $f_{a_1}^{\rm QCD}$, the scaling relations would imply
\begin{equation}\label{eq:frho_vs.mrho}
\frac{f^{\rm \,HC}_\rho}{m^{\rm \,HC}_\rho}
\sim 0.19\,, \qquad 
\frac{f^{\rm \,HC}_{a_1}}{ m^{\rm \,HC}_{a_1}} \sim 0.13\,,
\end{equation} 
in accord with Weinberg's sum rules that require $f^{\rm \,HC}_{a_1} /
m^{\rm \,HC}_{a_1} < f^{\rm \,HC}_\rho / m^{\rm \,HC}_\rho$.
Eq.~\eqref{eq:scaling} also implies
 \begin{equation}
 \frac{m^{\rm \,HC}_{a_1}} {m^{\rm \,HC}_{\rho }} \sim 1.6\,.
 \end{equation}
Motivated by \AFB, we consider $m_{\rho}^{\rm \,HC}
\sim 200$~GeV, corresponding to $f_\pi^{\rm \,HC} \sim 20 $~GeV. 

In the flavor symmetric limit, the flavor octet $\rho^a $ decays
primarily to pairs of HC pions, with the decay width
\begin{equation}
\label{eq:rhopipiwidth}
  \Gamma_{\rho\to\pi\pi }= {g_{\rho\pi\pi}^2\over32\pi}\,m_\rho\,\left(1- {4 m_\pi^2\over m_\rho^2}\right)^{\frac32}\,, 
\end{equation} 
where $g_{\rho \pi\pi}$ is the $\rho\,\pi \,\pi$ coupling in
\begin{equation}  
  {\cal L}_{\rho \pi \pi} = - g_{\rho \pi\pi} \,f_{a b c}\,\rho_\mu^a \,\pi^b \partial^\mu \pi^c\,.
\end{equation} 
The VMD estimate 
\begin{equation}\label{Eq:WMD:grhopipi}
g_{\rho \pi\pi}  \simeq m_\rho / f_\rho
\end{equation}
agrees with the NDA estimate $g_{\rho \pi \pi } \sim {\mathcal
  O}(4\pi)$ within a factor of $\sim 2$.  In QCD the VMD prediction is
only $16\%$ smaller than the measured value. Based on the above we can
expect $\Gamma_\rho/m_\rho\sim {\mathcal O}(10\%)$.

Flavor-symmetry breaking due to ${m_{\q}}_1 \ne {m_{\q}}_3$ splits the
$\rho\sim [\bar \q_{1,2}\q_{1,2}]$ and $K^*\sim [\bar \q_{1,2}\q_3]$
masses, as well as the corresponding $a_1$ and $K_1$ masses. In
general, $g_{VPP}~$(=$g_{\rho\pi\pi}$ in the flavor-symmetric limit)
now also depends on the vector and pseudoscalar flavors. Motivated by
QCD, we allow for ${\mathcal O}(10\%)$ flavor breaking.  The flavor
breaking also leads to ``${\omega}-\phi$" mixing as well as its
axial-vector analog. We denote the deviation from ideal vector-meson
mixing by the angle $\theta_V^{\rm id}$, so that the relation between
the mass eigenstates $V_{L,H}$ and the ideally mixed states
\begin{equation} \label{Vmixing}
  |V_{12}\rangle =  {1\over \sqrt{2}} (|\bar \q_1\,\q_1\rangle +|\bar \q_2\,\q_2 \rangle)\,,~~|V_{3}\rangle = |\bar \q_3\,\q_3 \rangle\,,
\end{equation}
is given by
\begin{equation}\label{Videal:def}
\begin{pmatrix}
|V_L\rangle\\|V_H\rangle
\end{pmatrix}
=
\begin{pmatrix}
\cos\theta_V^{\rm id}&\sin\theta_V^{\rm id}\\
-\sin\theta_V^{\rm id}&\cos\theta_V^{\rm id}
\end{pmatrix}
\begin{pmatrix}
|V_{12}\rangle\\ - |V_3\rangle
\end{pmatrix}\, ,
\end{equation}
and similarly for the axial-vector mixing angle $\theta_A^{\rm id}$,
with $V_{L,H}\to A_{L,H}$ and $V_{12,3}\to A_{12,3}$.  Frequently, we
will also employ QCD inspired notation for the mass eigenstates, i.e.,
$V_L=\omega_{\rm HC}$, $V_H=\phi_{\rm HC}$, and $A_L=f_1^{\rm HC}$,
$A_H=f_1'{}^{\rm HC}$. Since we do not include the ${}^{1}P_1$
resonances there is no equivalent of the $K_{1A}$-$K_{1B}$ mixing in
our simplified formalism.  In particular, we identify the $K_1^{\rm
  HC}$ in Table~\ref{tab:PG} with the equivalent of the $K_{1A}^{\rm
  QCD}$ in QCD.

Given $m_{\q_1}$ and $m_{\q_3}$, we determine the vector and
axial-vector masses as well as the mixing angles $\theta_{V,A}^{\rm
  id}$ using the simplified quark-model treatment of
Ref.~\cite{Cheng:2011fk}.  The HC hadronic parameters of this model
are obtained by fitting to the QCD vector and axial-vector meson data,
and then rescaling to the HC scale, $M_{\chi}$, as explained in detail
in Appendix~\ref{app:fit:to:spectra}.  The HC scale is defined to be
the $\rho$ mass in the chiral limit,
\begin{equation}
\label{eq:Mchi}
M_{\chi}\equiv \lim_{m_{{\cal Q}_{i}}\to 0} m_\rho.
\end{equation}
In turn, the $\pi$, $\rho$  and $a_1$ decay constants are taken to be
\begin{align}\label{eq:fpi}
f_\pi&=\frac{M_{\chi}}{m_\rho^{\rm QCD}}f_\pi^{\rm QCD},\\
 f_\rho&=\frac{M_{\chi}}{m_\rho^{\rm QCD}}f_\rho^{\rm QCD},\\
  f_{a_1}&=\frac{M_{\chi}}{m_{a_1}^{\rm QCD}}f_{a_1}^{\rm QCD}\label{eq:fa1}
\end{align}
where the QCD decay constants that we use are $f_\pi^{\rm
  QCD}=92$\,MeV, $f_{\rho}^{\rm QCD}=148$\,MeV, and $f_{a_1}^{\rm
  QCD}=168$\,MeV.

The $\rho_{\rm HC}$ resonances decay primarily to HC pion pairs, as in
QCD, and subdominantly to light-quark pairs,
cf. Table~\ref{tab:Br}. The $K^*_{\rm HC}$ resonances decay primarily
to $K_{\rm HC}+\pi_{\rm HC}$ pairs, as in QCD.  Their subdominant
decays to $t+$light quark pairs have a branching ratio of $\approx
30\%$, as explained in Section~\ref{sec:asymmetries}.  The fact that
the $K^*_{\rm HC}\!\to\! t+$jet decays are subleading is
phenomenologically favored and naturally achieved within our model, as
already mentioned in the Introduction. In our benchmark both the
$\omega_{\rm HC}$ and $\phi_{\rm HC}$ are kinematically forbidden to
decay to on-shell $\bar{K}_{\rm HC}K_{\rm HC}$ pairs.  Therefore,
their dominant decays are to SM quarks with very narrow decay widths,
cf. Table~\ref{tab:PG}.

For the axial-vector meson decay widths we use the model of
Ref.~\cite{Roca:2003uk}, where a global ${\rm SU}(3)$ flavor symmetry
is used for the matrix elements, but the phenomenologically more
important effect of flavor-symmetry breaking in the phase space of the
final states is kept. A hadronic parameter, $\tilde F_{\rm QCD}$,
obtained from the fit to the $A\to P V$ decay widths in
QCD~\cite{Roca:2003uk} is rescaled to $\tilde F_{\rm HC}$ in order to
obtain the corresponding HC decay widths.  See
Appendix~\ref{app:decay:widths} and Eq.~\eqref{eq:app:FHC} for
details. The HC $a_1$ state decays predominantly to $\rho_{\rm
  HC}\,\,\pi_{\rm HC}$ pairs, yielding a large decay width,
$\Gamma_{a_1}\simeq 0.2 m_{a_1}$. The branching ratio into light-quark
pairs is small, ${\mathcal O}({\rm few~} \%)$.  The HC $K_1$ resonance
decays predominantly to $K_{\rm HC}^* \,\,\pi_{\rm HC}$ pairs,
yielding $\Gamma_{K_1}\simeq 0.05 m_{K_1}$. The $K_1\to u,c+t$
branching ratio is of ${\mathcal O}(5\%-10\%)$. In our benchmark the
decays of the HC $f_1(=A_L)$ and $f_1'(=A_H)$ to $K_{\rm HC}
\,\,K^*_{\rm HC} $ pairs are kinematically forbidden.  Therefore, they
are very narrow with their dominant decays being to light-quark pairs.

\subsection{Would-be composite quarks\label{subsec:compquarks}}%
In our benchmark, an on-shell HC scalar ${\cal S}$ decays to 
$u_j \bar\q_j$ pairs well before
HC hadronization can occur.  In particular, the decay width of the heavy scalar
${\cal S}$ is
\begin{equation}\label{GammaS:part}
 \Gamma_{\cal S}=\sum_j \Gamma_{{\cal S} \to
  u_j\bar\q_j}\,, 
\end{equation}
where $j$ runs over $j=1,2,3$. The ${\cal S} \to u_j \bar\q_j$ partial
decay widths are given by
\begin{equation}
\begin{split}
\Gamma_{{\cal S} \to u_j \bar\q_j} \simeq {m_{\cal S} }\,
\frac{|h_j|^2}{16\pi}\,,
\end{split}
\label{eq:partonic:S}
\end{equation}
up to phase-space corrections which are at most of ${\mathcal
  O}(10\%)$.  In our benchmark the Yukawa couplings are large
($h_i\sim 2-4$), leading to $\Gamma_{\cal S}\simeq 0.44\times m_{\cal
  S}\simeq 230$\,GeV.  The ${\cal S}$ therefore decays on a timescale
that is much shorter than the HC hadronization timescale, which is
governed by $ \Lambda_{\rm HC}\!\sim\! {\mathcal O}({\rm
  few})f_\pi\!\sim\! {\mathcal O}(60{\rm~GeV})$.  Consequently, there
are no asymptotic bound states of two heavy HC scalars, ${\cal S}{\cal
  S}^*$, or of a HC scalar and a HC quark, ${\cal S}\q_i$. (Had we
taken the fundamental scalar to be much lighter, ${\cal S}{\cal S}^*$
bound states would form and clearly show up as resonances in the
differential $t \bar t$ spectrum. We are thus led to a consider scalar
mass $m_{\cal S} \gtrsim 0.5$\,TeV.)

However, the picture changes for production of the elementary quarks
$u_{Ri}$ via their Yukawa couplings to the composite operators ${\cal
  S}\q_i $.  The latter are also isosinglet QCD color triplets with
hypercharge $2/3$.  The SM right-handed up quarks can then be viewed
as an admixture of the elementary $u_{Ri}$ and bound-state ${\cal
  S}\q_i$ quark fields.  Note that in this case the scalar ${\cal S}$
has virtuality $\sqrt{s}\!\equiv( p_{\cal S}^2 )^{1/2}$ lying well
below $m_{\cal S}$.  The ${\cal S}$ decay width becomes $s$ dependent,
being obtained via the substitution $m_{\cal S}\to \sqrt{s}$ in
Eq.~\eqref{eq:partonic:S}, including the implicit phase-space factor.
Thus its decay width is suppressed to levels $\lsim \Lambda_{\rm HC}$,
and bound-state ${\cal S}\q_i$ quark fields with virtuality much
smaller than their would-be physical mass can form and mix into
on-shell $u_{Ri}$.

To estimate the mixing or partial compositeness of the $u_{Ri}$, we
assume that it is dominated by the lowest pole in the $T\{{\cal
  S}\q_i(x), {\cal S}^*\bar\q_i(0)\}$ two-point function, or
equivalently, by the lowest pole in the ${\cal S}\q_i\to {\cal S}\q_i$
scattering $S$ matrix. In the calculation of the mixing we will treat
the lowest pole as an asymptotic state.  Formally, this corresponds to
taking the limit $h_i \to 0$ in Eq.~\eqref{eq:partonic:S}, making
${\cal S}$ stable.

The Yukawa couplings induce mass mixing between the would-be composite
quarks and the elementary up quarks ($u_i=u,c,t$),
\begin{equation}\label{eq:u'mixing}
\sqrt{2} \, h_i  f_{u_i^\prime} \,
{\bar{u}}_{R i} \,{u^\prime}_{L i}\,,
\end{equation}
where the would-be composite quark decay constants $f_{u_i^\prime}$
are defined as
\begin{equation}\label{fu'def}
  \langle u_i^\prime  | \bar \q_{i}  \,{\cal S}^*  | 0\rangle
  =\sqrt{2}   f_{u_i^\prime} \,{\bar u}_i^\prime  \,. 
\end{equation}
Here the ${\bar u}_i^\prime$ are the Dirac spinors.

Since $m_{\q}\ll m_{\cal S}$ the would-be composite HC quarks
correspond to bound states of a heavy--light quark system. More
precisely, in our benchmarks $m_{\cal S}$ is $\sim 2 \Lambda_{\rm
  HC}$. Comparing to QCD this corresponds to a heavy--light quark
system with a heavy quark mass lying somewhere between the charm and
bottom quark mass. To estimate the $f_{u'_i}$ decay constants we
therefore interpolate between the known light--light and heavy--light
vector-meson decay constants in QCD, and rescale to the case of HC --
see Appendix~\ref{app:AMD}, Eq.~\eqref{eq:fu'} in particular, for
details.

For the purpose of this discussion we can take the ordinary $3 \times
3$ up-quark mass matrix to be flavor diagonal, neglecting the small
misalignment between the weak and up-quark mass bases in the SM. For
each generation the mixing between the SM and would-be HC quarks is
then described by $2\times 2$ matrices
\begin{equation} \label{eq:mass:matrix:ui}
  M_{RL}^ i= \left( 
  \begin{array}{ll} 
    m_{u_i} ~& \sqrt{2} h_i  f_{u_i^\prime}   \\
    0~& ~M_{u_i^\prime} 
  \end{array} 
  \right),~~i=1,2,3. 
\end{equation}
Here $m_{u_i} $ is the ordinary ${\rm SU}(2)_L$ breaking quark mass,
and $M_{u_i'}\simeq m_{\q_i}+m_{u_i}+ \Lambda_{\rm HC}$ is the mass
term for the would-be composite quark (see
Eq.~\eqref{eq:primemasses}). The mixing term follows from
Eq.~\eqref{eq:u'mixing}.

Diagonalization of Eq.~\eqref{eq:mass:matrix:ui} yields the mass
eigenstates
\begin{align}
\label{eq:masseigenstates}
  |u_{Ri }  \rangle^{\rm phys} &= \cos\theta_{R i } \,|u_{R i } \rangle -  \sin \theta_{R i } \,|u^\prime_{R i } \rangle\,, \\
  |u^\prime_{Ri  }  \rangle^{\rm phys} &= \sin\theta_{R i} \,|u_{R i } \rangle +  \cos \theta_{Ri } \,|u^\prime_{Ri } \rangle\,,
\end{align}
and similarly for the LH mass eigenstates with the replacement $R\to
L$.
The ordinary $u, c$, and $t$ quarks are identified with $u_1^{\rm phys}
$, $u_2^{\rm phys} $, and $u_3^{\rm phys} $, respectively.
Taking
$h_i f_{u'_i}$ significantly smaller than $M_{u'_i}$ yields
\begin{equation}
\label{eq:sintheta}
  \sin\theta_{R i} \sim \sqrt{2} h_i  {f_{u_i^\prime} \over
    M_{u_i^\prime}}\,,~~~\sin\theta_{Li} \sim \sqrt{2} h_i
            {f_{u_i^\prime} m_{u_i}  \over M^2_{u_i^\prime} }\,. 
\end{equation}
The RH mixings are substantially larger than the LH ones, which are
suppressed by the $M_{u_i'}$. Specifically, for our benchmark we find
\begin{align}
\label{eq:mixinganglesRH}
 & \sin\theta_{R1} = \sin\theta_{R2} = 0.22\,, & \sin\theta_{R3} = 0.43\,, \\
\label{eq:mixinganglesLH}
 & \sin\theta_{L1} = \sin\theta_{L2} \approx 0\,, & \sin\theta_{L3} = 0.10\,.
\end{align}

The couplings of the vector mesons to the would-be composite quarks
are given by
\begin{equation}
\label{eq:rhouprime}
  {\cal L} = g_{\rho  } (\bar{u}^\prime T^a \gamma^\mu u^\prime)
  \rho^a_\mu + g_{a_1 } (\bar{u}^\prime T^a \gamma^\mu \gamma_5
  u^\prime)  {a_1}^a_\mu \,+\cdots \,.
\end{equation}
Here, the $u'$ appear in the interaction basis of
Eqs.~\eqref{eq:u'mixing}--\eqref{eq:mass:matrix:ui} and, for
simplicity, we have taken flavor-symmetric couplings. The ellipses
denote higher-derivative operators.  In NDA, both $g_{\rho}$ and
$g_{a_1 }$ are ${\mathcal O}(4\pi )$, while the VMD estimates are (see
Appendix~\ref{app:AMD})
\begin{equation}  
  \label{eq:VMDgrhoga1} 
  g_{\rho }  \simeq {m_\rho \over  f_\rho}\,,~~~g_{a_1 }\simeq {m_{a_1} \over  f_{a_1}}\,, 
\end{equation}
so that $g_\rho\simeq g_{\rho \pi \pi}$.  In the numerics below we will take them to be equal. 

Couplings of the vector mesons to the SM quarks are induced 
then via the quark mixing in Eq.~\eqref{eq:masseigenstates}.
In the quark-mass basis these couplings are given by
\begin{equation}~\label{eq:rhoqq}
\begin{split}
  {\cal L} = & \lambda^R_{ij}\, \bar{u}_{Ri} \gamma^\mu   T_{ij}^a  \rho^a_\mu u_{Rj}   
	      + (R \to L ) +\cdots \,,
\end{split}
\end{equation}
where 
\begin{equation}
\label{eq:lambdai}
 \begin{split}
    \lambda^R_{ij}=&~ g_{\rho }\sin\theta_{Ri}\sin\theta_{Rj}\,,
  \end{split}
\end{equation}
and the ellipses denote terms involving the would-be composite quarks. 

The LH quark couplings $\lambda_i^L$ are obtained by substituting
$R\to L$ in Eqs.~\eqref{eq:rhoqq}--\eqref{eq:lambdai}.  The
axial-vector meson couplings to quarks follow by substituting
$\gamma^\mu \to \gamma^\mu \gamma_5$ and $g_{\rho } \to g_{a_1}$.  The
$K^*ut$ coupling $\lambda_{113}^R$ and the corresponding $K_1 ut$
coupling are phenomenologically important for NP contributions to
$t\bar t$ production from t-channel HC resonance exchanges. On the
other hand, the s-channel contributions from $\phi/\omega$ as well as
$f_1/f_1'$ exchanges are suppressed by the small $\theta_{V,A}^{\rm
  id}$ mixing angles.  The above couplings also govern the partial
decay widths of the vector and axial-vector resonances to quark pairs.

Similarly, the interactions of the HC pions with the SM quarks follow
from their couplings to the would-be composite quarks,
\begin{equation}
{\cal L} = \frac{g_A}{f_\pi} \Big(\bar{u}^\prime_{Ri}
\, T^a_{ij} \, \cancel{\partial} \pi^a u^\prime_{Rj} - \bar{u}^\prime_{Li}
\, T^a_{ij} \, \cancel{\partial} \pi^a u^\prime_{Lj}\Big)\,+\cdots,
\end{equation}
where the ellipses again denote higher-derivative operators.  In NDA
$g_A\sim {\mathcal O}(1)$, consistent with the QCD nucleon-pion axial
coupling $g_A^{\rm QCD}\simeq 1.26$.  Ignoring the spin structure, as
warranted in the heavy-scalar limit, one can also compare $g_A$ with
the QCD $B^*B\pi$, and $D^*D\pi$ couplings, which are also ${\mathcal
  O}(1)$ (i.e., $\hat g\sim0.6-0.7$ \cite{Abada:2003un, Ohki:2008py,
  ElBennich:2010ha}).

Changing to the physical quark basis, the Lagrangian is given by
\begin{equation}\label{eq:pionL}
\begin{split}
{\cal L} = 
 \frac{g_A}{f_\pi}&  \sin\theta_{Ri}\sin\theta_{Rj} \, \bar{u}_{Ri}\,
 T^a_{ij} \, \cancel{\partial} \pi^a u_{Rj} - \big(R \to L\big)\,,
\end{split}
\end{equation}
where we do not show the terms that involve would-be HC quarks.
Integrating by parts and using the Dirac equation,
Eq.~\eqref{eq:pionL} is equivalent on the quark mass shell to
\begin{equation}
\begin{split}\label{eq:pion:couplings}
  {\cal L} \simeq &  
\frac{ig_A}{f_{\pi}} \sin\theta_{Ri}\sin\theta_{Rj} m_{u_j} \,
\bar{u}_{Ri}\,  T_{ij}^a \pi^a u_{Lj} \\
&+\rm h.c. - \big(R \to L\big)\,.
\end{split}
\end{equation}

We see that the only significant contribution is proportional to
$m_t$. Thus, production of $t\bar t$ pairs receives important
contributions from t-channel $K$ exchange. In contrast, s-channel
contributions are suppressed by the light-quark masses.

The couplings in Eq.~\eqref{eq:pion:couplings} are also responsible
for pion decays to quark pairs, e.g. $\pi\to 2j$, $K\to jt^*$.  The
$\pi^a$ decay widths are given by
\begin{equation}
  \begin{split}\label{eq:pi:dec}
    &\frac{\Gamma_{\pi^a \to \bar{u}_{i} u_{j}}}{m_{\pi^a}} = \frac{ g_A^2 N_c}{16\pi} 
    \frac{ \big( m_{u_j}^2 + m_{u_i}^2 \big)}{f_{\pi}^2}    T^a_{ij} T^a_{ji}
  \sin\theta_{Ri}^2 \sin\theta_{Rj}^2  ,
  \end{split}
\end{equation}
where $a=1,2,3$, and we do not write down terms further supressed by
the light-quark masses. While the decay widths are narrow due to
light-quark mass suppression, they do not lead to displaced vertices
since they correspond to decay lengths of tens of nanometers.

\section{Tevatron and LHC Phenomenology\label{sec:benchmarks}}%
Next, we assess the effect of the new HC sector on the $t\bar t$
production cross sections and asymmetries at the Tevatron and the LHC.
The relevant measurements and the corresponding SM predictions are
collected in Table~\ref{tab:exp} and Table~\ref{tab:sm}.  As an
example we take the benchmark set of parameters introduced in the
previous section.  It has been chosen to demonstrate that our model
can easily yield anomalously large $A_{\rm FB}^{t\bar t}$ asymmetries,
while satisfying the remaining $t\bar t$ constraints. We also show
that the benchmark passes the Tevatron and LHC dijet tests that often
invalidate models addressing the Tevatron $A_{\rm FB}^{t\bar t}$
anomalies. Cross sections for the production of new resonances that
are present in our model are evaluated, and the most promising signals
are identified.  In Section~\ref{sec:EWK} we will discuss another
class of constraints, namely electroweak precision tests, including
atomic-parity violation.

\subsection{The top-antitop asymmetries: experimental review \label{sec:asymmetries}}%
The CDF and D\O\, experiments at the Tevatron have measured various
partonic level asymmetries in $p\bar p\to t\bar t$.  One of these is
the inclusive asymmetry,
\begin{equation}
A_{\rm FB} \equiv \frac{N(\Delta y>0) - N(\Delta y<0)}{N(\Delta y>0) + N(\Delta y<0)}\,,
\end{equation}
where $\Delta y = y_t-y_{\bar t}$ is the difference between the $t$
and $\bar t$ rapidities, taking the forward direction to be that of
the proton. The SM prediction for the inclusive asymmetry is
\begin{equation}
A_{\rm FB}^{\rm SM} = 0.088\pm 0.006\,.
\label{eq:pred:AFB}
\end{equation}
This result uses NLO cross-section differences, including leading EW
corrections, in the numerator \cite{Kuhn:1998jr, Kuhn:1998kw,
  Kuhn:2011ri, Ahrens:2011uf, Manohar:2012rs, Hollik:2011ps,
  Bernreuther:2012sx}; and LO cross sections in the denominator.  Note
that use of the NLO cross sections in the denominator would reduce the
predicted asymmetry by ${\mathcal O}(30\%)$.  The NLO PDF set {\tt
  CTEQ6.6m} \cite{Nadolsky:2008zw} is used throughout. The error in
Eq.~\eqref{eq:pred:AFB} is the pure scale uncertainty for
$\mu\in[m_t/2,2m_t]$.

The CDF \cite{Aaltonen:2012it} measurement of $A_{\rm FB}$ is larger
than the SM prediction, as was the 2011 D\O\, measurement.  A new
D\O\, measurement \cite{Abazov:2014cca}, extended to include a 3-jet
sample in $t\bar t$ production, is significantly lower and supersedes
the previous one.  Naively averaging with CDF yields
\begin{equation}
A_{\rm FB}^{\rm inc}=0.124\pm0.025\,.
\end{equation}

\begin{table}
\centering
\begin{tabular}{lll}
\hline\hline
{\bf Observable}				&{\bf Value}		&{\bf Ref.}\\
\hline
$A_{\rm FB}^{\rm low,\,  CDF}$ 			& $0.084\pm0.055$	&\cite{Aaltonen:2012it}	\\		
$A_{\rm FB}^{\rm high,\, CDF}$ 			& $0.295\pm0.067$ 	&\cite{Aaltonen:2012it}	\\ 		
$A_{\rm FB}^{\rm inc,\,  CDF}$ 			& $0.164\pm0.047$ 	&\cite{Aaltonen:2012it}	\\	
$A_{\rm FB}^{\rm inc,\,  {\rm D\O}}$ 			& $0.106\pm0.030$ 	&\cite{Abazov:2014cca}  	\\	
$A_{\rm FB}^{\rm inc,\,  average}$		& $0.124\pm0.025$ 	&			\\[1em]	
$\AC^{\rm inc,\,ATLAS,\, semileptonic}$ 	& $0.006\pm0.010$	&\cite{ATLAS-CONF-2013-078}	\\
$\AC^{\rm inc,\,ATLAS,\, dileptons}$   	& $0.057\pm0.028$	&\cite{ATLAS:2012sla}		\\
$\AC^{\rm inc,\,CMS,\, semileptonic}$   	& $0.004\pm 0.015$   	&\cite{Chatrchyan:2012cxa}	\\
$\AC^{\rm inc,\,CMS,\, dileptons}$     	& $-0.010\pm0.019$      	&\cite{Chatrchyan:2014yta}	\\
$\AC^{\rm inc,7TeV,average}$                 	& $0.007\pm0.008$    	&				\\[1em]
$\AC^{\rm inc,CMS,8TeV}$                 	& $0.005 \pm 0.009$    	&\cite{CMS-PAS-TOP-12-033}	\\[1em]
$\sigma_{\rm inc}^{\rm CDF+D\O}$        		& $(7.60 \pm 0.41)\pb$    	&\cite{Aaltonen:2013wca}	\\
$\sigma_{\rm inc}^{\rm ATLAS}$ (7 TeV)		& $(177 \pm 11) \pb$		&\cite{ATLAS:2012fja}		\\
$\sigma_{\rm inc}^{\rm ATLAS}$ (8 TeV)		& $(237.7  \pm 11.3) \pb$	&\cite{ATLAS-CONF-2013-097}	\\
$\sigma_{\rm inc}^{\rm CMS}$   (7 TeV)		& $(165.8 \pm 13.3) \pb$	&\cite{CMS-PAS-TOP-11-024}	\\
$\sigma_{\rm inc}^{\rm CMS}$   (8 TeV)		& $(239 \pm 13) \pb$	&\cite{Chatrchyan:2013faa}	\\[1em]
\hline\hline
\end{tabular}
\caption{Experimental input 
for $t \bar t$ production cross section and asymmetries. All errors
have been added in quadrature.}
\label{tab:exp}
\end{table}

\begin{table}
\centering
\begin{tabular}{lll}
\hline\hline
{\bf Observable}				&{\bf Value}		&{\bf Ref.}\\
\hline
$A_{\rm FB}^\text{low, SM}$                   & $0.062\pm0.003$                       &\cite{Bernreuther:2012sx}\\
$A_{\rm FB}^\text{high, SM}$                  & $0.129\pm0.006$                       &\cite{Bernreuther:2012sx}\\
$A_{\rm FB}^\text{inc, SM}$                       & $0.088\pm 0.006$                      &\cite{Bernreuther:2012sx}\\[1em] 
$\AC^{\rm SM}$ (7 TeV)                           & $0.0123 \pm 0.0005$                   &\cite{Bernreuther:2012sx}\\
$\AC^{\rm SM}$ (8 TeV)                           & $0.0111 \pm 0.0004$                   &\cite{Bernreuther:2012sx}\\[1em] 
$\sigma_{\rm inc}^{\rm TEV,\,NNLO}$		& $(7.395 \pm 0.544 )\pb$		&\cite{Czakon:2013tha}\\
$\sigma_{\rm inc}^{\rm LHC,\,NNLO}$ (7 TeV)	& $(172.5 \pm 15.0) \pb$		&\cite{Czakon:2013tha}\\
$\sigma_{\rm inc}^{\rm LHC,\,NNLO}$ (8 TeV)	& $\big(246.3 ^{+19.8}_{-20.5}\big) \pb$	&\cite{Czakon:2013tha}\\
\hline\hline
\end{tabular}
\caption{SM predictions
for $t \bar t$ production cross section and asymmetries. 
 }
\label{tab:sm}
\end{table}

Interestingly, CDF observes a significant rise in $A_{\rm FB}$ with
the invariant mass of the $t\bar t$ pair. Quoting just their two-bin
result as an example, they report
\begin{equation}
A_{\rm FB}^{\rm low}=0.084\pm0.046\pm0.030\,,
\end{equation}
 for $m_{t\bar t} \leq 450$\,GeV and  
\begin{equation}
A_{\rm FB}^{\rm high}=0.295\pm0.058\pm0.033\,,
\end{equation}
 for $m_{t\bar t} \geq 450$\,GeV, which should to be
compared to the SM predictions~\cite{Bernreuther:2012sx}
\begin{equation}
A_{\rm FB}^{\rm low, SM}=0.062\pm0.003\,,
\end{equation}
and 
\begin{equation}
A_{\rm FB}^{\rm high,SM}=0.129\pm0.006\,.
\end{equation}
The CDF and D\O\, collaborations have also presented results with
finer, albeit different, binning in $m_{t\bar  t}$~\cite{Abazov:2014cca,Aaltonen:2012it}. 
The two sets of measurements are consistent with each other, with the exception of the 
largest bin, $m_{t\bar t} > 650$\,GeV, for which D\O\, obtains a negative central value 
with an error that is 68\% larger than CDF's. In the $m_{t\bar t} \in [550,650]$ bin, 
the D\O\, and CDF central values are very close but the D\O\, error is 60\% larger.
The CDF fitted slope for $A_{\rm FB}$ vs $m_{t\bar t}$ is $1.8\sigma$ larger than D\O's, 
and $2.4\sigma$ larger than the NLO SM prediction. Both collaborations have also 
measured $A_{\rm FB}$ vs.~the rapidity difference $\Delta y$, again with different binning.
The CDF fitted slope is $1.3\sigma$ larger than D\O's, and $2.4\sigma$ larger than the NLO SM prediction.

At the LHC, the initial state is symmetric, thus there is no fixed
forward or backward direction with respect to which an asymmetry can be defined.  Instead, the 
observable that is related to $A_{\rm FB}$ is the charge asymmetry,
\begin{equation}
\AC = \frac{N(\Delta|y|>0)-N(\Delta|y|<0)}{N(\Delta|y|>0)+N(\Delta|y|<0)}\,,
\label{eq:acdef}
\end{equation}
where $\Delta |y|=|y_t|-|y_{\bar t}|$ is the difference between the absolute
values of the top and antitop rapidities.  At $7$\,TeV both ATLAS
and CMS have measured the charge asymmetry in the semileptonic and
dilepton decay channels, albeit with appreciable experimental
uncertainties (see Table~\ref{tab:exp}). Naively averaging the four
measurements yields
\begin{align}
\AC^{\rm EXP} &= 0.007 \pm 0.008\,,
\label{}
\intertext{consistent with the SM prediction \cite{Bernreuther:2012sx}}
\AC^{\rm SM}     &=  0.0123 \pm 0.0005\,.
\label{}
\end{align}
(Only averaging the two semileptonic measurements yields $\AC^{\rm
  exp} = 0.005 \pm 0.009$.)  Again, the SM prediction has been
obtained with the LO cross section in the denominator.  An 8 TeV
measurement of $\AC$ was recently presented by
CMS~\cite{CMS-PAS-TOP-12-033},
\begin{equation}
\AC=0.005\pm0.009\,,
\end{equation}
which is also consistent with the SM prediction~\cite{Bernreuther:2012sx}
\begin{equation}
\AC=0.0111\pm0.0004\,.
\end{equation}

Whether or not the experimental situation at the Tevatron points to an
anomalously large forward-backward asymmetry or is due to statistical
fluctuations, our philosophy will be to show that in our model large
enhancements of $A_{\rm FB}$ can be consistent with all other
constraints, thus highlighting the stealth nature of the new strong
interactions.

\subsection{Choosing a benchmark  \label{sec:benchmark}}%
We calculate the asymmetries in our NP model using the procedure
outlined in Ref.~\cite{Drobnak:2012rb}, and employed in
Ref.~\cite{Bernreuther:2012sx} for the SM predictions given above. In
the numerator we take the SM at NLO (QCD + EW), and work with LO cross
sections in the denominators. The contributions from NP (including
NP--SM interference) in both the numerator and denominator are always
evaluated at LO.  All LO cross sections are automatically evaluated in
{\tt Madgraph} \cite{Alwall:2011uj} using the NLO PDF-set {\tt CTEQ6m}
with a fixed renormalisation and $\alpha_s$ scale. For the benchmark
presented here we fix the renormalisation scale to $\mu=2m_t$.

To obtain the Tevatron and LHC total $t\bar t$-production cross
sections and differential $d\sigma/dm_{t\bar t}$ spectra we use the
NNLO {\tt CTEQ10} predictions \cite{Czakon:2013tha} at $\mu=m_t$ for
the total SM cross sections, with their reported errors (see
Table~\ref{tab:sm} ); and {\tt aMC@NLO} for the differential SM
spectra, evaluated at $\mu=m_t$, with the errors reflecting the scale
and pdf uncertainties. The NP contributions to the total and
differential spectra are evaluated at LO for a fixed scale choice,
$\mu=2 m_t$, as in the asymmetries.

Two of the explanations that have been proposed for the potential
Tevatron $A_{\rm FB}^{t\bar t}$ anomalies are t-channel exchange of
light vectors, e.g., $W'$, $Z'$ \cite{Jung:2009jz}, or of light
scalars \cite{Blum:2011fa}. In both cases the exchanged particle's
mass is optimally a few hundred GeV or less. The models can yield a
large $A_{\rm FB}^{t\bar t}$ that, particularly in the case of vector
exchange, increases appreciably with $m_{t\bar t}$. Moreover, both
proposals have been shown to simultaneously lead to good agreement
with the $d\sigma/dm_{t\bar t}$ spectra (for the t-channel exchanges,
correcting for the CDF acceptance at large pseudorapidity is crucial
\cite{Gresham:2011pa, Jung:2011zv, Grinstein:2011dz}).

Our model provides a concrete renormalizable example that combines the
two proposals. The role of the $Z'$ is played by the HC $K^*$ and, to
a lesser extent, by the $K_{\rm 1}$. The t-channel scalar corresponds
to the HC $K$.  Note that for high $m_{t\bar t}$ there are also
perturbative contributions coupling to the RH up-type quarks from
intermediate $S - \q $ box graphs, which scale as
\begin{equation}
 {h_1^2 h_3^2 \over 16 \pi^2 } {1\over m_{t\bar t}^2 } \sim   {1\over 2
   m_{t\bar t}^2 }\,.
\end{equation}
However, we find that their effects are subleading and do not consider
them further.

At the LHC, important constraints come from $\sigma_{t\bar t}$ and
$\AC$. An increase in $A_{\rm FB}^{t\bar t}$ via t-channel exchange is
typically correlated with an increase in $\AC$ beyond its measured
value. However, associated light-mediator production, e.g. $gq\to
t+(Z'\to \bar tq)$, has been shown to reduce $\AC$
\cite{Drobnak:2012rb}.  Associated light-mediator production also
contributes to the total LHC cross section, $\sigma_{t\bar t}$. The
resulting constraint, as well as the ATLAS and CMS bounds from $t$+jet
resonance searches, are evaded if the light mediator has other open
decay channels, thus suppressing the $Z'\to \bar tq$ branching ratio.
In our model a new dominant decay channel is naturally present. In
particular, the strong interaction decay $K^* \to K\pi$ can lead to
${\rm Br}(K^*\to \bar t j)\sim{\mathcal O}(30\%)$. This would still
allow for a significant reduction of $\AC$.  Note that on-shell $K\to
t+j$ decays are kinematically forbidden, so that the above constraints
do not apply.

\begin{figure}
  \centering
  \includegraphics[width=0.45\columnwidth]{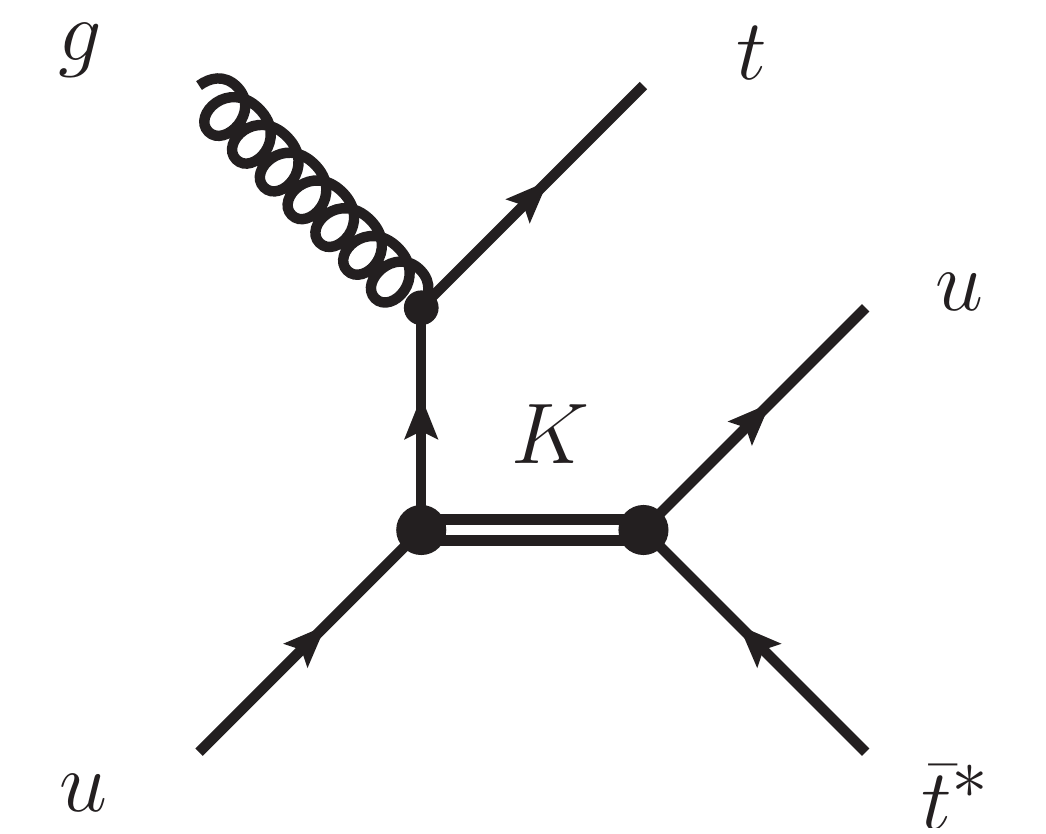}  
  \includegraphics[width=0.45\columnwidth]{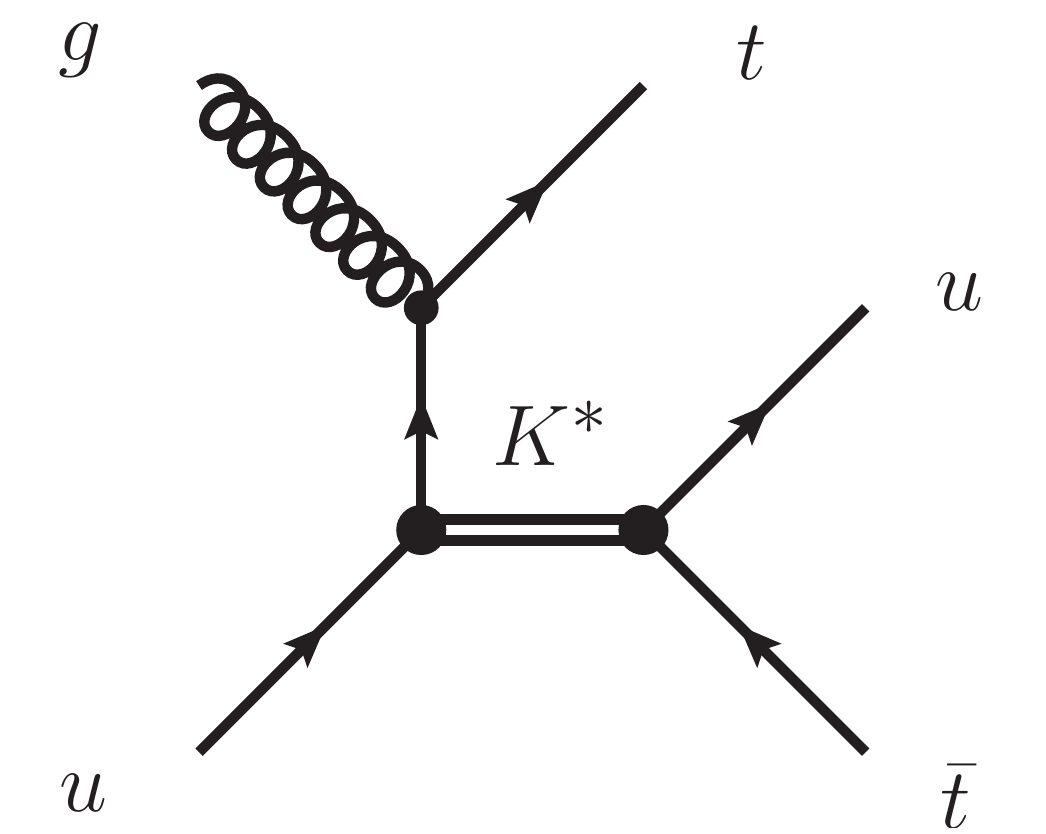}  
  \caption{Feynman diagrams for the associated production of a $K$ and
    $K^*$, respectively. The $\bar t$ resulting from the
   $K$ decay is off-shell.}
  \label{fig:assoc-K}
\end{figure}

Previous studies, as outlined above, thus motivate us to search for
benchmarks with relatively light $K^*$ and $K$, with masses of
$\sim200$\,GeV. Moreover, a $\rho$ mass in this range is also favored
by the recent CDF bounds on pair production of dijets
\cite{Aaltonen:2013hya}.  For $\rho\to \pi\pi\to jj\,jj$ with
$m_\rho\lesssim 200$\,GeV and $m_\pi\sim$ 70 GeV the bounds weaken
significantly and, in fact, lie above the expected limits.

In order to obtain a viable set of parameters for our model that i)
yields substantially enhanced $A_{\rm FB}^{\rm t\bar t}$ at the
Tevatron and ii) yields agreement with all other constraints, we
employ a rough $\chi^2$-minimization procedure containing a subset of
available measurements. Minimizing the $\chi^2$ with respect to a
large number of variables is algorithmically difficult. We use the
{\tt COBYLA} method \cite{ANU:1770520}, which allows us to apply
constraints on the minimization procedure.

The $\chi^2$ contains the experimental values of the inclusive $A_{\rm
  FB}^{t\bar t}$ and $\AC$ ($7$\,TeV), the total $t\bar t$ cross
sections at the Tevatron and the LHC ($7$\,TeV), and the highest bins
of the differential cross sections at ATLAS ($7$\,TeV), and CMS
($8$\,TeV). The UV inputs are $h_1, h_3, m_{{\cal Q}_1}, m_{{\cal
    Q}_3}$, $m_{\cal S}$, and the HC scale $M_{\chi}$ defined in
Eq.~(\ref{eq:Mchi}). Through dimensional transmutation, the latter is
equivalent to the choice of HC strong coupling constant, $\alpha_{\rm
  HC}$, in the UV.  The UV parameters fix the pseudoscalar masses via
the quadratic terms in the HC chiral Lagrangian, and the vector and
axial-vector masses via the naive quark model described in
Appendix~\ref{app:fit:to:spectra}.  The would-be composite quark
masses in the interaction basis are given by
Eq.~\eqref{eq:primemasses}.  The decay constants $f_\pi, f_\rho$, and
$f_{a_1}$ (for simplicity taken to be universal for all members of the
corresponding flavor octets), are given in
Eqs.~\eqref{eq:fpi}--\eqref{eq:fa1}. The would-be composite quark
decay constants $f_{u'_i}$ are allowed to vary within $30\%$ of the
interpolation given in Eq.~\eqref{eq:fu'}.

\begin{figure*}[]
\begin{center}
\hspace*{-1em}\includegraphics{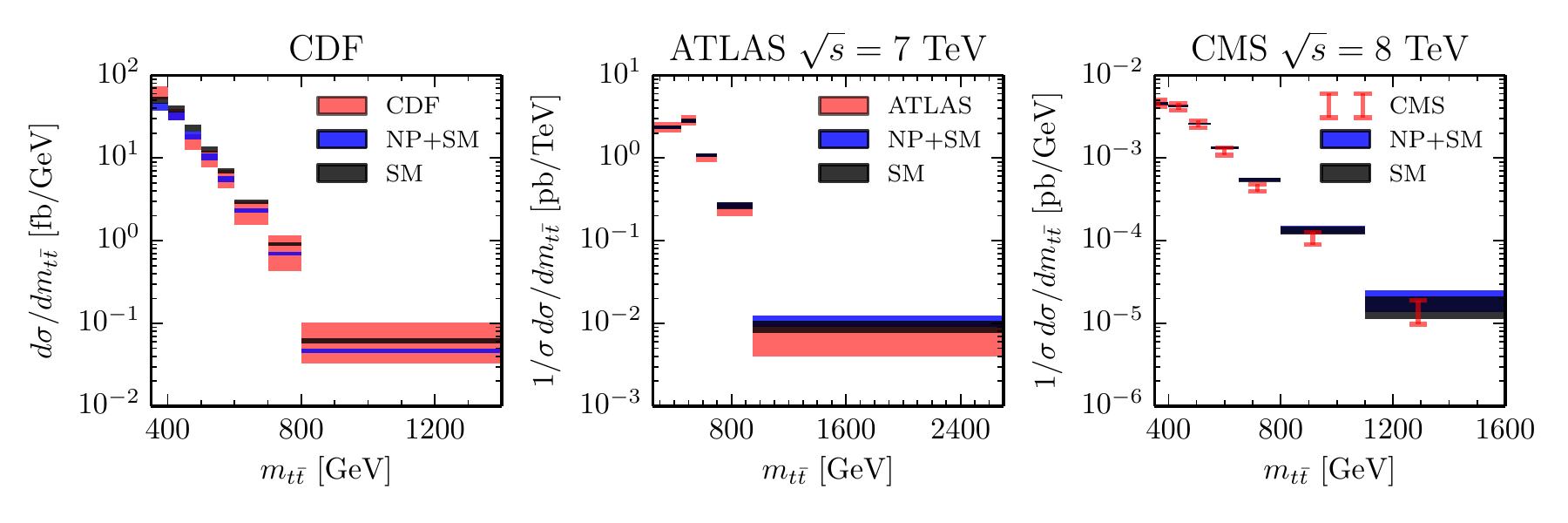}
\end{center}
\caption{The differential cross sections, $d\sigma/dm_{t\bar t}$, at
  Tevatron (first panel) and at the LHC ($7$\,TeV ATLAS in the second and
  $8$\,TeV  CMS in the third panel), SM prediction in black, the NP
  benchmark predictions in blue and measurements by CDF given by red
  bands. } 
\label{fig:diff:xsec}
\end{figure*}

\begin{figure}[]
\begin{center}
\hspace*{-1em}\includegraphics[]{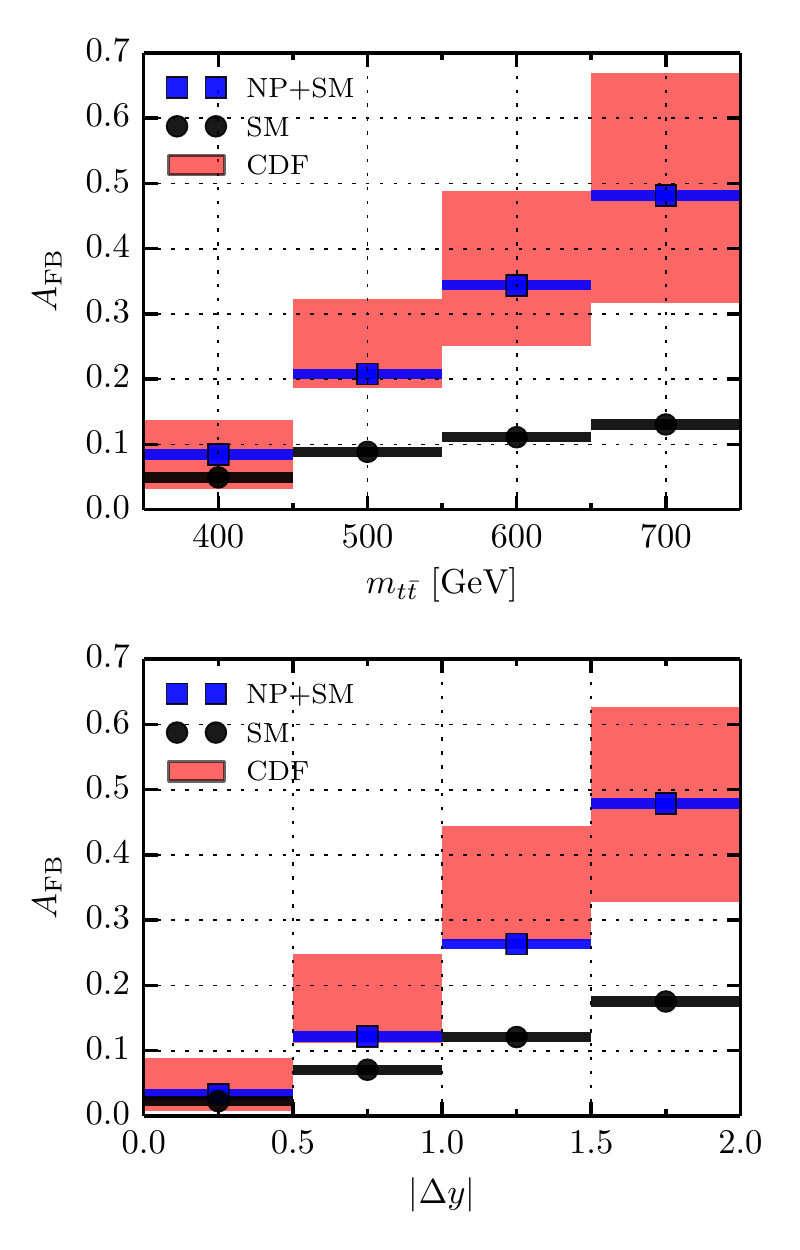}
\end{center}
\caption{Differential $A_{\rm FB}$ asymmetries at the Tevatron as a
  function of $m_{t\bar t}$ (top) and $\Delta y$ (bottom) with the SM
  prediction in black, the NP benchmark predictions in blue and the
  measurements by CDF given by red bands.}
\label{fig:differential:asymmetries}
\end{figure}

We must also choose values for the couplings of the HC resonances to
the would-be composite quarks, i.e. $g_\rho$, $g_{a_1}$, and $g_A$.
We allow $g_\rho$ and $g_{a_1}$ to lie within roughly $30\%$ of the
values obtained from Eq.~\eqref{eq:VMDgrhoga1}, and we take $g_A
=1.26$, identifying it with the representative QCD nucleon-pion axial
coupling.  The vector-pion coupling $g_{\rho\pi\pi}$ is taken to be
equal to $g_\rho$. The decay widths of the vector meson multiplet,
$\rho, K^*, ...$, are determined via
Eqs.~\eqref{vector:decay:widths}--\eqref{eq:mixed:dec:width}.  The
axial decay widths are determined using the model of
\cite{Roca:2003uk}, see Eqs.~\eqref{eq:GAVP}--\eqref{eq:app:FHC}, with
the parameter $\tilde F_{\rm HC}$ fixed to the value obtained from
Eq.~\eqref{eq:app:FHC}.  The pseudoscalar decay widths
\eqref{eq:pi:dec} are small and do not enter into our analysis.

\begin{table}
\begin{ruledtabular}
\begin{tabular}{ccc}
  HC Resonance & channel & Br(\%)\\\hline
  $\rho_{\rm HC}$ &  $\pi\pi$ & $98.2$  \\
  &  $\bar uc, \bar cu, \bar u u + \bar c c$ & $1.8$  \\[0.5em]
  $K^*_{\rm HC}$ & $K \pi$ & $68$  \\
  & $\bar ut, \bar tu, \bar ct, \bar tc$ & $32$  \\[0.5em]
  $V_H[\phi_{\rm HC}]$ &  $\bar u u + \bar c c$ & $100$  \\
  $V_L[\omega_{\rm HC}]$ &  $\bar u u + \bar c c$ & $100$ \\
\end{tabular}
\end{ruledtabular}
\caption{Table of the dominant branching ratios of HC vector
  resonances and their decays into SM quarks.}
\label{tab:Br}
\end{table}

The UV or fundamental parameters for our illustrative benchmark are
listed in Table~\ref{tab:bmuv}, together with the resonance couplings
and decay constants. The resonance masses and decay widths are given
in Table~\ref{tab:PG}.  Realization of the phenomenologically favored
range ${\rm Br}(K^*\to \bar t j)\sim 30\%$ arises via phase-space
suppression of the dominant $K^*\to K\pi$ decay mode, see
Table~\ref{tab:Br}. Since $m_K^*\simeq m_K+m_\pi$ the phase-space
factor is small, of $O(10^{-2} )$ in our benchmark. The tuning
associated with the phase-space suppression is actually quite
moderate, given that the approximate equality of $m_K^*$ and
$m_K+m_\pi$ changes relatively slowly as the HC quark masses
$m_{\q_3}$, $m_{\q_1}$ are varied. For instance, the Barbieri-Giudice
measure of fine tuning for the phase-space suppression factor,
corresponding to variation of the HC quark masses around the benchmark
point and using the naive quark model for the vector masses, is
$\approx 8$.  It is comparable to the tuning associated with the
coincidence of $m_\phi$ and $2 m_K$ in QCD.

Before moving to the resulting phenomenology, we comment on the large
benchmark Yukawa couplings $h_3 =4.2 $, $h_1 = 2.0$.  This is driven
by two factors: a sizable product of couplings $h_1 h_3 $ is required
in order to obtain a large t-channel enhancement of the
forward-backward asymmetry, e.g. $A_{\rm FB}> 0.15$; and $h_1 $ is
bounded from above by dijet constraints, most notably the CDF bounds
on dijet pair production, see below.  A moderate decrease in $h_3$ is
possible if the non-perturbative couplings $g_\rho $ or $g_A$ of the
vector or pseudoscalar mesons to the would-be composite quarks are
moderately increased, or if $h_1$ is maximized consistently with dijet
phenomenology.  Nevertheless, a large $h_3$ is required, e.g., $h_3 >
3$.
 
The values of the Yukawa couplings $h_1$, $h_3$ in
Table~\ref{tab:bmuv} correspond to a renormalization scale which can
be approximately identified with $m_{\cal S}$.  A large $h_3 $ at this
scale prompts us to ask if our theory is sensible at higher energies.
For example, whether we encounter a Landau pole as we evolve $h_1$,
$h_3$ and the HC gauge coupling $g_{\rm HC}$ upwards in energy. To
answer this, we fix the values of $h_1 (m_{\cal S})$, $h_3 (m_{\cal
  S})$ to those given in Table~\ref{tab:bmuv}.  We also take $g_{\rm
  HC} (m_{\cal S} ) =1.9$, the value we would obtain for the QCD
coupling $g_s$ at scale $\mu =m_{\cal S}\, f_\pi /f_\pi^{\rm HC}$ by
running upwards with 3 flavors rather than 4 (given the 3 HC quark
flavors) from its usual two-loop $\overline{\rm MS}$ value at $\mu
=1$\,GeV.  The one-loop RGEs yield a Landau pole at $\mu \approx
1.6$\,TeV.  However, moving to two-loop RGEs using the general results
from Ref.~\cite{Luo:2002ti} we find that $h_3$ reaches an approximate
attractive UV fixed point at $ \mu = O(10) $ TeV, given by $h_3^* =
16\, \pi /\sqrt{37}\approx 8.3$, while $h_1$ and $g_{\rm TC}$ are
asymptotically free.  For lower values of $h_3 (m_{\cal S} ) > 3$ (see
above), with $h_3 (m_{\cal S}) > h_1 (m_{\cal S})$, the one-loop
Landau pole and the two-loop UV fixed point for $h_3$ are,
respectively, reached at scales that are a few times larger.  Clearly,
given the large value of $h_3^* $ obtained at two-loops, the question
of whether a true UV fixed-point exists or not can only be settled
using nonperturbative methods.

Consistency of our model requires that there are no QCD and HC
breaking condensates, $\langle \bar u_{R3} \, \q_{L3} \rangle \ne 0$
and $\langle {\cal S}\rangle\ne0$, which could potentially be
triggered by a large value of $h_3$.  An estimate of the critical
Yukawa coupling above which condensates form can be obtained using the
Schwinger-Dyson equation at one-loop in the rainbow or ladder
approximation in the massless scalar limit (see~
Ref.~\cite{Hung:2010xh}, where such an estimate was applied to
electroweak-symmetry breaking via fourth-family condensates with large
Higgs Yukawa couplings). In our model, the ladder approximation in the
$m_{\cal S} =0$ limit yields $h_3^{\rm crit} = 2 \pi$, somewhat below
the two-loop fixed point coupling $h_3^*$.  If this result also holds
nonperturbatively, the field content of the theory would need to be
enlarged in order for the model to be phenomenologically viable. For
example, we have checked that the addition of massive singlets, ${\cal
  N}_i$, with flavor conserving yukawa couplings to the HC quarks,
\begin{equation}
 h_{{\cal N}1}^i {\cal N}_i (\bar \q_1 \q_1 + \bar \q_2 \q_2 ) +
h_{{\cal N}3}^i {\cal N}_i \bar \q_3\q_3\,, 
\end{equation}
lead to an asymptotically free $h_3$, with $h_3$ always well below
$2\pi$, for a large set of $h_{{\cal N}1,2}^i$ values. In this case
some or all of the singlet yukawas, $h_{{\cal N}1,2}^i$, obtain a
fixed point. The presence of the singlets also has an added benefit
that the HC quark masses, $m_\q$, can be generated dynamically via the
induced singlet vevs.

\subsection{Top-antitop asymmetries and cross sections: benchmark predictions \label{sec:asymmetries}}%
The predictions for the Tevatron $t\bar t$ asymmetries within our
benchmark are (quoting the central values)
\begin{equation}
A_{\rm FB}^{\rm inc}=0.173\,, \quad A_{\rm FB}^{\rm low}=0.091\,,
\quad A_{\rm FB}^{\rm high}=0.301\,,
\end{equation}
corresponding to a large enhancement of $A_{\rm FB}$ at large
$m_{t\bar t }$.  On the other hand, the charge asymmetries at the LHC
are predicted to be
\begin{equation}
\AC^{\rm inc, 7{\rm TeV}}=0.0137\,, \quad \AC^{\rm inc, 8TeV}=0.0135\,,
\end{equation}
consistent with the SM predictions, as well as their measured values.
Note that the associated production of $K^*$ has a significant effect
on the value of $\AC$. Without this effect the charge asymmetries
would have been $\AC^{\rm inc, 7{\rm TeV}}=0.0245$, and $\AC^{\rm inc,
  8{\rm TeV}}=0.0239$. The total cross sections at the Tevatron and
the LHC are found to be
\begin{equation}
\begin{split}
\sigma^{\rm TEV}_{\rm inc} =6.34&\pm0.54{\rm pb}\,, \\
\sigma^{\rm LHC}_{\rm inc} \text{($7$\,TeV)} =176&\pm 15{\rm pb}\,, \\ 
\sigma^{\rm LHC}_{\rm inc} \text{($8$\,TeV)} = 252&\pm 20{\rm pb}\,,
\end{split}
\end{equation}
where the errors reflect the uncertainty in the SM contributions at
NNLO, as discussed above. These predictions are in good agreement with
the experimental measurements, listed in Table~\ref{tab:exp}, with the
exception of a $\sim 2 \sigma $ tension with the larger measured value
of $\sigma^{\rm TEV}_{inc}$.  Note that the NP contributions to all
observables have been treated at leading order and are therefore
subject to significant uncertainties which have not been included in
our predictions.

The differential forward-backward asymmetries $dA_{\rm FB}/dm_{t\bar
  t}$ and $dA_{\rm FB}/d|{\Delta y}|$ are compared to the CDF
data\footnote{Since our main point is to provide an explicit model
  which can explain a large asymmetry while being consistent with all
  other data, we compare our predictions to the CDF measurements,
  which yield larger slopes than D\O's.} and the SM predictions in
Figure~\ref{fig:differential:asymmetries}. The CDF differential cross
section is shown in Figure~\ref{fig:diff:xsec} (left).  The dominant
NP effect on $t\bar t$ production in our model is due to t-channel
exchanges.  Thus, the effect of the CDF rapidity acceptance
corrections for large $m_{t\bar t}$ is significant~\cite{Jung:2011zv,
  Gresham:2011pa}. We take this into account using the prescription in
Ref.~\cite{Grinstein:2011dz}.  In Figure~\ref{fig:diff:xsec} we
compare the predicted normalized differential cross section, $1/\sigma
\,d\sigma/dm_{t\bar t}$, with the 7 TeV ATLAS~\cite{Aad:2012hg} and 8
TeV CMS~\cite{CMS:fxa} measurements for semileptonic final states.  We
can see that it is not difficult to reproduce the increase in $A_{\rm
  FB}$ vs.~$m_{t\bar t}$ and $A_{\rm FB}$ vs.~$\Delta y$.  A modest
indication of the well known high $m_{t\bar t}$ tail in the LHC
$d\sigma/dm_{t\bar t}$ distribution, characteristic of low-scale
t-channel exchanges, can be seen in the last bin of the second as well
as the third panels of Figure~\ref{fig:diff:xsec}.  It lies well
within the experimental uncertainties.

The deficit in the inclusive $t\bar t$ cross section at the Tevatron,
$\sigma^{\rm TEV}_{\rm inc}$, is primarily due to the lowest bin, as
can be seen in Figure~\ref{fig:diff:xsec} (first panel). Note that a
relative increase in the scalar ($K$) vs.~vector ($K^*, K_1$)
contributions to $t\bar t $ production would reduce this deficit.
This could be achieved by increasing the coupling $g_A$ relative to
$g_{\rho, a_1}$.

\begin{table}
\begin{ruledtabular}
\begin{tabular}{ccc}
  HC Resonance & channel & Br(\%)\\\hline
  $\pi_{\rm HC}$ &  $\bar uc, \bar cu, \bar u u + \bar c c$ & $100$ \\[0.5em]
  $K_{\rm HC}$ & $\bar ut, \bar tu, \bar ct, \bar tc$ & $100$  \\[0.5em]
  $\eta_{\rm HC}$ &  $\bar u u + \bar c c$ & $100$ \\
\end{tabular}
\end{ruledtabular}
\caption{Table of the dominant branching ratios of HC pions into SM
  quarks.}
\label{tab:Brpions}
\end{table}

\begin{table}
\begin{ruledtabular}
\begin{tabular}{ccc}
  HC Resonance & channel & Br(\%)\\\hline
  $a_{1}$ &  $\rho\pi$ & $99.08$  \\
  &  $\bar uc$, $\bar cu$, $\bar u u + \bar c c$ & $0.92$  \\[0.5em]
  $K_{1}$ & $\rho K$ & $92.6$  \\
  & $\bar ut, \bar tu, \bar ct, \bar tc$ & $7.4$  \\[0.5em]
  $A_L$ & $\bar u u + \bar c c$  & $100$  \\[0.5em]
  $A_H$ &  $\bar u u + \bar c c$ & $100$ \\
\end{tabular}
\end{ruledtabular}
\caption{Table of the dominant branching ratios for HC axial-vector
  resonances and their decays to the SM quarks.}
\label{tab:Br:a1}
\end{table}

\subsection{Dijets\label{sed:dijets}}%
The dijet cross-section measurements at the Tevatron and the LHC
typically provide stringent constraints on models that aim to explain
the forward-backward asymmetry in $t\bar t$, since the resonances are
usually required to have large couplings to quarks. The s-channel
exchanges are subject to direct resonance searches (i.e. bump hunting
in $pp\to 2j$), while t-channel exchanges could visibly enhance the
$d\sigma_{jj}/dm_{jj}$ spectra at large invariant
masses~\cite{Grinstein:2011dz}.

The couplings of the various resonances to light-quark pairs in our
benchmark are summarized in Table~\ref{tab:quark-couplings}.  Dijet
production in the s-channel is primarily due to $\rho$, $\omega$, and
$a_1$ exchanges. The $\rho$, $\omega$ and $a_1$ contributions are
suppressed by their relatively small couplings to light quarks. This
is a result of the hierarchy between $h_1$ and $h_3$, see
Table~\ref{tab:bmuv}.  Moreover, the $\rho$ and $a_1$ contributions
are further suppressed by their small branching ratios to quark pairs
(they predominantly decay to PP and VP pairs, respectively,
cf. Table~\ref{tab:Br} and \ref{tab:Br:a1}).  Finally, the s-channel
contributions of the pseudoscalars are negligible because of the
chiral suppression of their couplings to light quarks.

\begin{table}
\begin{ruledtabular}
\begin{tabular}{cccc}
HC Resonance & quarks &$\kappa^V_{R}$ & $\kappa^V_{L}$\\
\hline
$\rho$ & $\bar uu, \bar cc$  & $\pm$ 0.117$$ &$ 0.0$\\
& $ \bar u c$ &$ 0.165$ &$ 0.0$\\[0.5em]
$K^*$ &$ \bar u t, \bar c t$ &$ 0.328$ &$ 0.0$\\[0.5em]
$V_L$& $ \bar u u, \bar c c$ &$ 0.117$ &$ 0.0$\\
& $\bar t t$ & $-0.018$ & $-0.001$\\[0.5em]
$V_H$&$ \bar u u, \bar c c$ & $-0.003$ &$ 0.0$\\
& $\bar t t$ & $-0.649$ & $-0.038$\\
\hline
$a_1$ & $\bar uu, \bar cc$  & $\pm0.161$ &$ 0.0$\\
& $ \bar u c$ &$ 0.228$ &$ 0.0$\\[0.5em]
$K_1$ &$ \bar u t$,$ \bar c t$ &$ 0.451$ &$ 0.0$\\[0.5em]
$f_1$& $ \bar u u, \bar c c$ &$ 0.160$ &$ 0.0$\\
& $\bar t t$ & $-0.116$ & $-0.007$\\[0.5em]
$f_1'$&$ \bar u u, \bar c c$ & $-0.021$ &$ 0.0$\\
& $\bar t t$ & $-0.887$ & $-0.052$
\end{tabular}
\end{ruledtabular}
\caption{HC resonance couplings to SM quarks. They
  correspond to the coefficients in the Lagrangian of Eq.~\eqref{eq:rhoqq}
  after rotating all fields to the mass eigenbasis. }
\label{tab:quark-couplings}
\end{table}

All of the above resonances also contribute in the t-channel. Here the
branching ratios to dijets play no role, since the resonance
contributions only depend on their couplings to the light quarks.  The
modest hierarchy $h_1< h_3$ in Eq.~\eqref{eq:handm} turns out to be
crucial.  For instance, had we taken $h_1\simeq h_3$ the t-channel
exchanges would yield an appreciable ${\mathcal O}(1)$ excess at
$m_{jj}=3$\,TeV.

\begin{figure}[]
\begin{center}
\hspace{-1em}\includegraphics[]{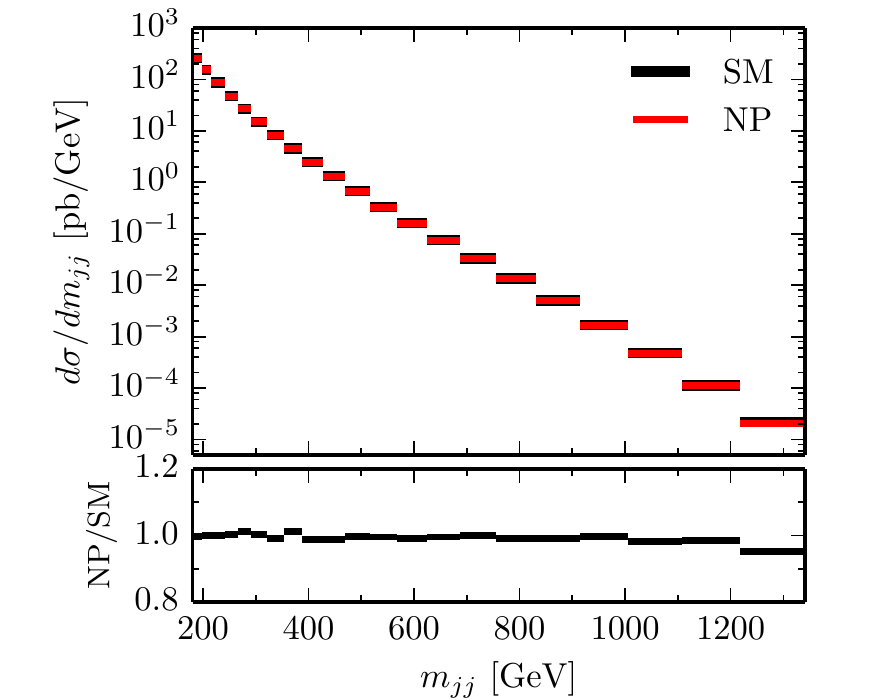}
\end{center}
\caption{The dijet cross-section distribution at CDF.}
\label{fig:tev:dijetspectrum}
\end{figure}

\begin{figure}[]
\begin{center}
\hspace*{-1em}\includegraphics[]{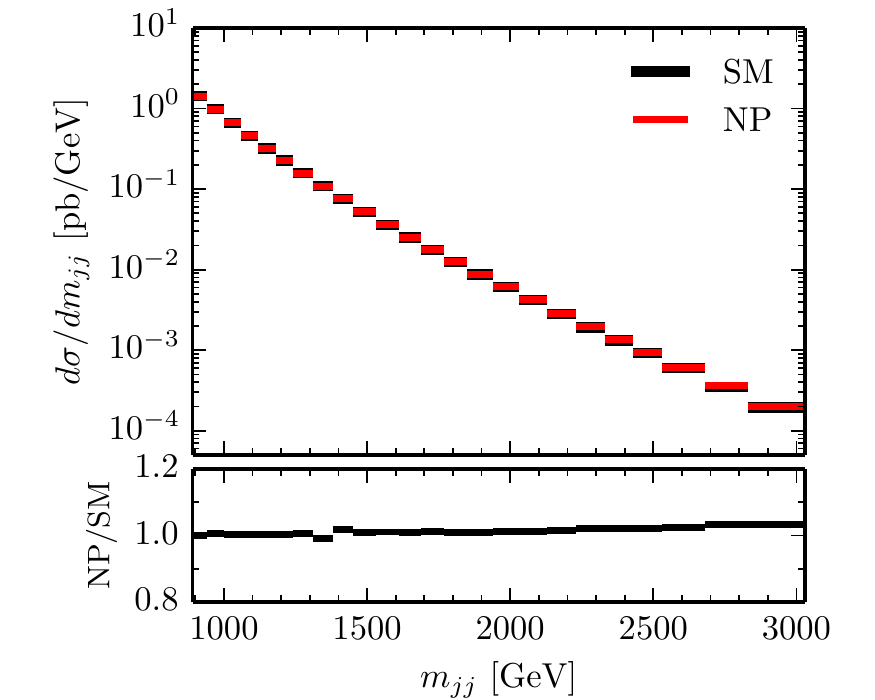}
\end{center}
\caption{The dijet cross-section distribution at CMS.}
\label{fig:dijet:CMS}
\end{figure}

In Figures~\ref{fig:tev:dijetspectrum} and \ref{fig:dijet:CMS} we
compare the benchmark and SM $d\sigma_{jj}/dm_{jj}$ dijet mass spectra
at the Tevatron and LHC (8 TeV). The dijet cross sections are
calculated at the partonic level at LO, using ${\tt MadGraph}$ with
${\tt CTEQ6M}$ and NLO $\alpha_s$.  Guided by the experimental
analyses~\cite{Aaltonen:2008dn, Chatrchyan:2013qha} we impose the
following cuts on the two outgoing partons (i.e. the two leading
jets). For the Tevatron we impose $|y|<1$. For the 8 TeV LHC cross
section calculation we require that the pseudorapidity difference
between the two partons satisfies $\Delta \eta_{jj}<1.3$, and that
$|\eta|<2.5$, $p_T>30$~GeV for each of them. The renormalization scale
is set to the average $p_T$ of the outgoing partons in both cases.  In
the LHC analysis the dijet mass is above $m_{jj}>890$~GeV.

The upper two panels in Figures~\ref{fig:tev:dijetspectrum}
and~\ref{fig:dijet:CMS} show $d\sigma_{jj}/dm_{jj}$ in the SM (black
line) and in our benchmark (red line).  The lower two panels show the
ratios of the two,
$(d\sigma_{jj}^{NP}/dm_{jj})/(d\sigma_{jj}^{SM}/dm_{jj})$.  The effect
of the new resonance exchanges is small, lying below the experimental
uncertainties at both the Tevatron and the LHC.  In both cases the
experimental analysis was aimed at bounding resonance production in
the dijet channel. The CDF bump hunting analysis allows for about a
$1\%$--$2\%$ spread in the ratio of data to a smooth background for
$m_{jj}\in [200,700]$\,GeV.  This spread is larger than the deviation
of $(d\sigma_{jj}^{NP}/dm_{jj})/(d\sigma_{jj}^{SM}/dm_{jj})$ from 1,
as shown in the lower panel of
Figure~\ref{fig:tev:dijetspectrum}. Furthermore, our benchmark does
not show any bumps in the spectrum at this level of precision. The CMS
bump-hunting analysis allows for a NP contribution in the $m_{jj}$
spectrum at the level of a few per mill at $m_{jj}\sim 1000$\,GeV,
with an increase to ${\mathcal O}(10\%)$ at $m_{jj}\sim 3000$\,GeV.
Note that the benchmarks differential distribution is very smooth.
Fitting $d\sigma_{jj}^{NP}/dm_{jj}$ to the same analytical function
that was used to describe the smooth QCD background in
Ref.~\cite{Chatrchyan:2013qha}, we find that the difference between
the fit and the prediction is always well below a per mill. Thus, the
bump-hunting analysis is not sensitive to our model.

\begin{figure}[]
\begin{center}
\hspace*{-1em}\includegraphics[]{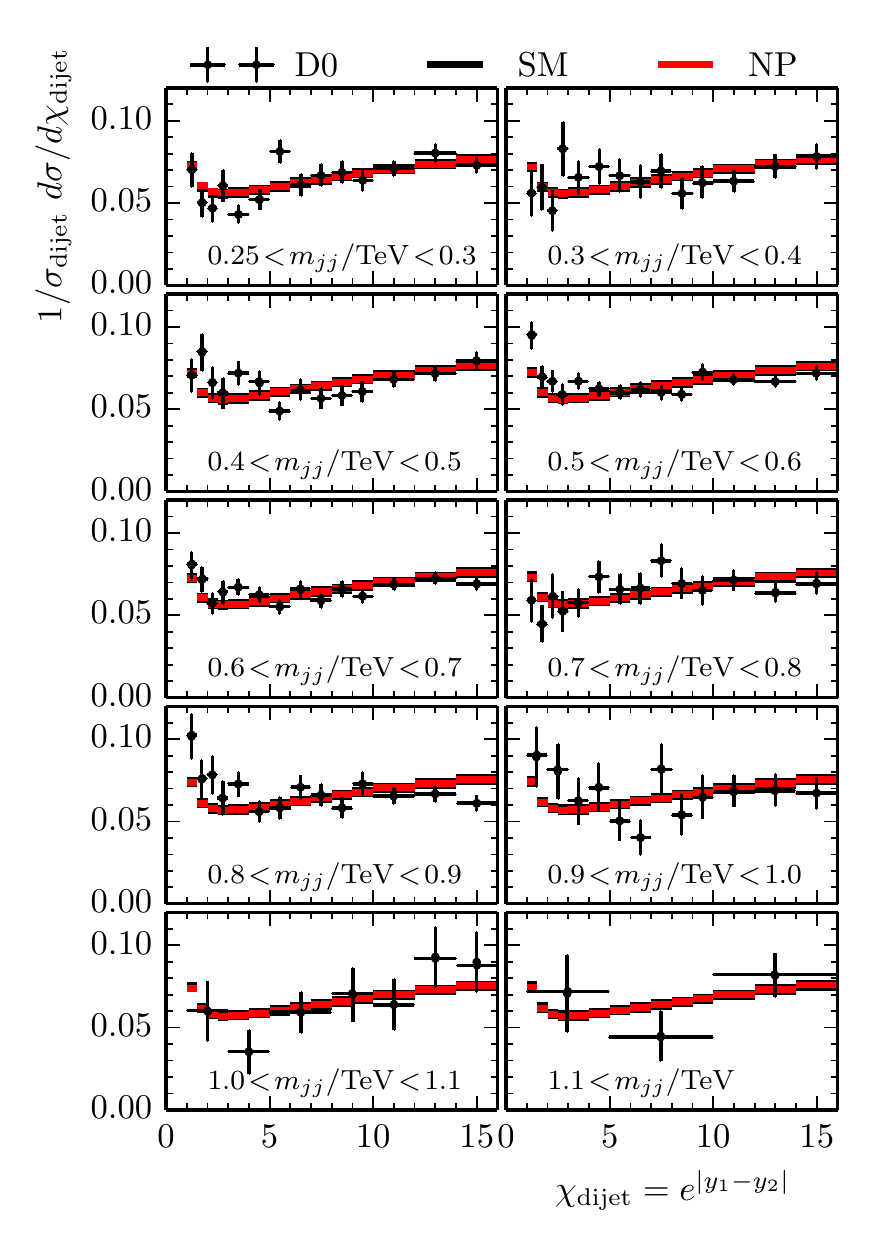}
\end{center}
\caption{The dijet angular distributions at the Tevatron, in bins of
  $m_{jj}$.  The SM predictions are denoted by black lines, the
  benchmark predictions in red, while measurements are denoted with
  crosses of the size of error bars.}\label{fig:angular:Tevatron}
\end{figure}

\begin{figure}[]
\begin{center}
\hspace*{-1em}\includegraphics[]{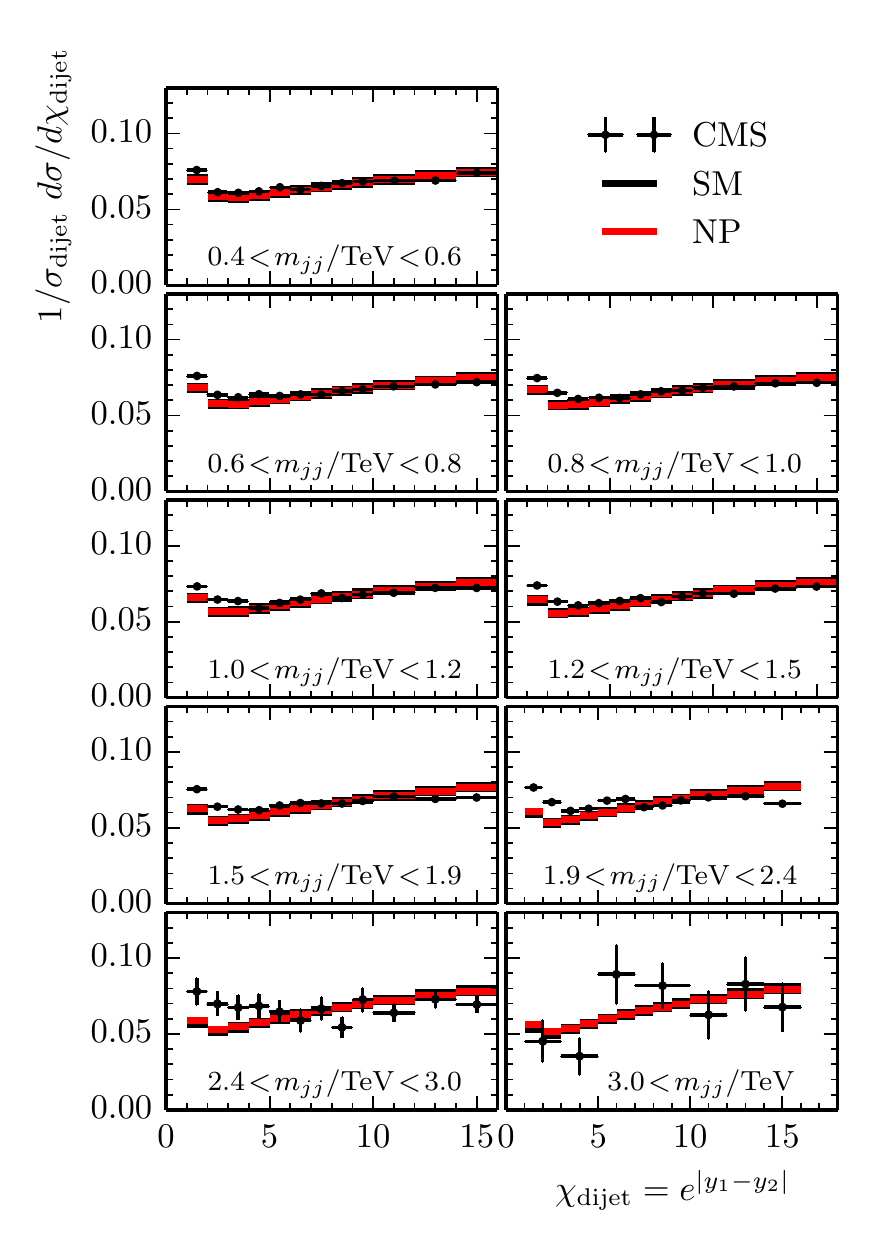}
\end{center}
\caption{The angular distributions in dijet production at the LHC,
  measured in bins of $m_{jj}$ as indicated in the plots. The SM
  prediction is denoted by black lines, the prediction with our
  benchmark NP in red, while crosses denote the measured spectra
  including errors.}\label{fig:angular:LHC}
\end{figure}

CMS and D\O\, have also measured the dijet angular distributions
$d\sigma/d\chi$ as functions of the dijet mass (here
$\chi=\exp(|y_1-y_2|)$, where $y_1$ and $y_2$ are the rapidities of
the two leading jets)~\cite{Abazov:2009ac, Chatrchyan:2012bf}. The
comparison of our benchmark and SM predictions are shown in
Figure~\ref{fig:angular:Tevatron} for the Tevatron and in
Figure~\ref{fig:angular:LHC} for the LHC. The predictions are
calculated at LO at the partonic level using {\tt Madgraph} with {\tt
  CTEQ6M} pdfs, setting the renormalization scale to the average $p_T$
of the outgoing partons.  Following the D\O\, analysis, we impose the
Tevatron cut $y_{\rm boost}\equiv0.5|y_1+y_2|<1$, where $y_{1,2}$ are
now the rapidities of the two partons (as opposed to the rapidities of
the two leading jets). The D\O\, measurements begin at
$m_{jj}>250$~GeV.  Following the CMS angular analysis, we impose the
LHC cut $y_{\rm boost}<1.11$.  The CMS measurements begin at
$m_{jj}>400$\,GeV. The contributions from the NP resonances lead to
deviations from the SM predictions that are much smaller than the
experimental error bars.  In the figures we show the LO predictions
for the SM, however the NLO predictions are
available~\cite{Nagy:2001fj, Nagy:2003tz}.  They further improve the
agreement between the data and the SM predictions.  Our conclusion
that the NP contributions to the angular distributions are negligible
is not expected to change when going from LO to NLO predictions.

Another constraint arises from searches for pair production of
resonances that decay to dijets, resulting in 4-jet final states. In
our model this signal would be due to s-channel $\rho$ production
followed by $\rho\to\pi\pi$ decays with $\pi\to jj$. The 95\% CL bound
on $\sigma(p\bar p\to X\to YY\to jj\; jj)$ from CDF for $m_X=175$~GeV
and $m_Y=50(70)$~GeV is $66.8(111.5)$~pb~\cite{Aaltonen:2013hya}. In
our benchmark, $X=\rho$ with a mass of $177$~GeV and $Y=\pi$ with a
mass of $62$~GeV.  The inclusive production cross section at LO is
$\approx 79$~pb.  However, after imposing partonic cuts based on the
CDF hadronic jet cuts ($p_T^\text{min}>15$\,GeV and $|\eta|<2.4$) we
obtain $\sigma(p\bar p \to \rho \to \pi\pi \to jj\,jj)=37\,$\,pb.  The
$O(\alpha_s )$ $Z'$ production $K$-factor, $1 + 8 \pi \alpha_s (\mu)
/9 $, increases this cross section by $\approx 30\% $ (at $\mu =
m_\rho = 177$\,GeV) to 48 pb.  The analyses of pair production of
dijets at CMS and ATLAS probe $m_{jj}>250$\,GeV and
$m_{jj}>150$\,GeV~\cite{Chatrchyan:2013izb, ATLAS:2012ds},
respectively, and are thus not sensitive to the $pp\to \rho\to\pi\pi$
mode in our model.  However, they could be relevant for production of
higher resonances, which we cover in the next section.

\subsection{Production of new states\label{new-res}}%
In this section we discuss existing constraints on the production of
HC resonances in our model.  As already mentioned, the
CDF~\cite{Aaltonen:2012qn}, CMS \cite{Chatrchyan:2012su} and
ATLAS~\cite{ATLAS-CONF-2012-096} collaborations have searched for
$t+j$ resonances, which could, in principle, constrain associated
$K^*t$ and $K_1 t$ production. The CDF and ATLAS analyses put bounds
on $t+j$ resonance masses above $m_{tj}>200$~\,GeV, and are thus
relevant for our model (the CMS obtains bounds for
$m_{tj}>400$\,GeV). The associated $K^*t$ and $K_1 t$ cross sections
are listed in Table~\ref{tab:qprime}. Here we sum over the CP
conjugate final states $K^*t$ and $\bar K^* \bar t$ as well as over
the light flavors, $K_{13}^*\sim[Q_1 \bar Q_3]$ and $K_{23}^*\sim[Q_2
  \bar Q_3]$, and similarly for the $K_1$.  At the 7\,TeV LHC one has
$\sigma_{K^*t}\,Br_{K^*\to \bar t j}=4.4$\,pb, which is roughly a
factor of 5 below the ATLAS bound for
$m_{K^*}=211$\,GeV~\cite{ATLAS-CONF-2012-096}. At the Tevatron
$\sigma_{K^*t}Br_{K^*\to \bar t j}=0.07$\,pb, which is roughly an
order of magnitude smaller than the CDF bound \cite{Aaltonen:2012qn}.
In the case of associated $K_1$ production the products $\sigma_{K_1t}
\, Br_{K_1\to \bar tj}$ lie even further below the corresponding
bounds at the Tevatron and the LHC.

Associated $Kt$ production leads to a $\bar t^* t j$ final state, with
one of the top quarks off-shell.  This feeds into the experimental
measurements of the (inclusive) $t\bar t$ cross sections
\cite{Chatrchyan:2013faa, ATLAS-CONF-2013-097, CMS-PAS-TOP-12-006} and
the $Wt$ production cross section \cite{Chatrchyan:2012zca,
  ATLAS-CONF-2013-100}. The $Kt$ cross section is comparable to the
theory error on the SM prediction for $t\bar t$
production. Furthermore, since the $t^*$ is off-shell, only a fraction
of the $Kt$ signal spills over into the $t\bar t$ production
cross-section measurements. For instance, using a LO {\tt Madgraph}
analysis and imposing the experimental cuts for the $t\bar t$ signal
region employed in the recent CMS dileptonic analysis
\cite{Chatrchyan:2013faa}, we estimate the $Kt$ contribution to the
$8$\,TeV $t\bar t$ cross section to be below $11$\,pb. It is thus
smaller than the error on the measurement $\sigma(p p \to t\bar
t)=239\pm13$\,pb \cite{Chatrchyan:2013faa}. The softer $Kt$ lepton
$p_T$ significantly reduces the leakage into the signal
region. Similarly, the $Kt$ contribution to the $Wt$ production signal
region in the recent CMS dilepton analysis at 7 TeV is below
$1.7$\,pb, to be compared with the CMS measurement of $\sigma(p p\to
Wt)=16^{+5}_{-4}$\,pb \cite{Chatrchyan:2012zca}.

\begin{table}
  \begin{ruledtabular}
  \begin{tabular}{lllll}
Final state &$\sigma_{\rm TEV}$ & $\sigma_{\rm LHC7}$ & $\sigma_{\rm LHC8}$ & $\sigma_{\rm  LHC13}$\\
\hline
$K t$ & 0.38  & 18.0 & 24.2& 64.5\\
$K^{*} t$ & 0.22 & 13.6  & 18.5 & 50.6 \\
$K_1 t$ & 0.11 & 11.1  & 15.4 & 45.1 \\
  \end{tabular}
  \end{ruledtabular}
  \caption{Inclusive cross sections for $pp \to K_\text{HC}^{(*)} t$
    and $pp \to K_1^{(*)} t$ associated production at the Tevatron,
    LHC 7TeV, 8TeV and 13TeV (in pb). Summation over the first two
    generations $\q_{1,2}$ and the CP-conjugate modes is assumed.}
  \label{tab:qprime}
\end{table}

\begin{table}
\begin{ruledtabular}
\begin{tabular}{llll}
& [7 TeV] & [8 TeV] & [13 TeV] \\
\hline
 $\sigma$ (pb)  & 0.54 & 0.92 & 4.39 \\
\end{tabular}
\end{ruledtabular}
\caption{Cross sections for LHC pair production of the
  HC scalars ${\cal S}$ ${\cal S}^*$, at various center-of-mass
  energies.}
\label{tab:qprime:pair}
\end{table}

Next we move to pair production of the colored scalars, ${\cal S}{\cal
  S}^*$. As discussed in Section~\ref{subsec:compquarks} the decay
width of the ${\cal S}$ is almost an order of magnitude greater than
the HC hadronization scale $\Lambda_{\rm HC}\sim {\mathcal O}({\rm
  few}) f_\pi$. Therefore the ${\cal S}$ scalars decay before they can
hadronize. This is reminiscent of the top quark in QCD. The ${\cal S}$
decays to quark--HC-quark pairs, ${\cal S}\to u_i\bar \q_i$, where
$i=1,2,3$. The ${\bar \q}_i$ from ${\cal S}\to u_i\bar \q_i $ and the
$\q_j$ from ${\cal S}^*\to u_j\bar \q_j$ hadronize via HC strong
interactions and result in final states containing many $\pi_{\rm
  HC}$, $K_{\rm HC}$.  In general we expect $pp\to {\cal S}{\cal
  S}^*\to u_i \bar u_j X_{\bar \q_i \q_j}$, where $X_{\bar \q_i \q_j}$
is the multi-$\pi$, $K$ state.  The light $\pi$ will receive sizable
boosts.  Thus, a $\pi \to jj$ decay will on average appear as a single
``fat jet'' in the detector.  However, ``fat jets'' from the heavier
$K \to t^* + j$ decays should be easier to resolve.

\begin{figure}[]
\begin{center}
\hspace*{-1em}\includegraphics[]{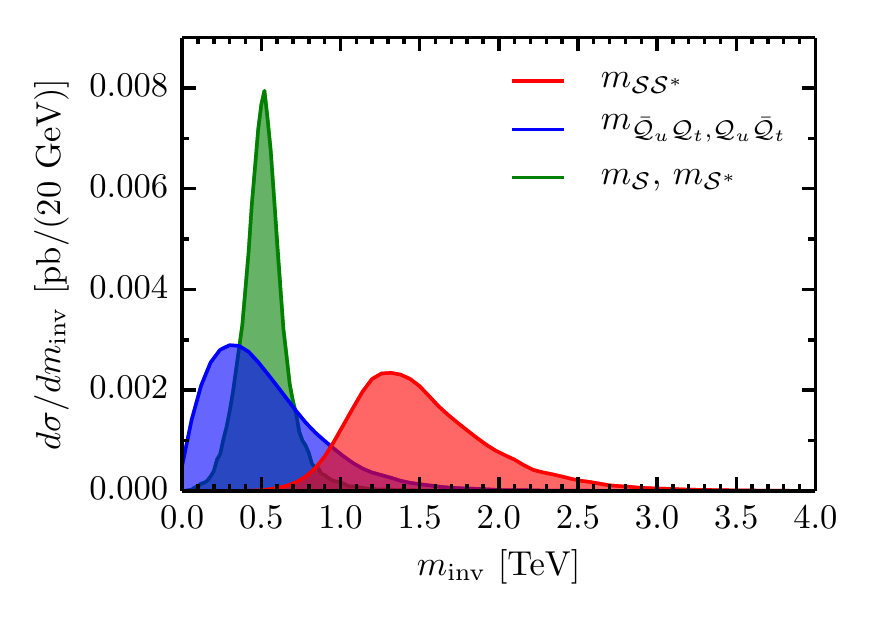}
\end{center}
\caption{The invariant mass distributions for $pp\to {\cal S}{\cal
    S}^*\to q_{1,2} \bar t X_{\bar \q_{1,2} \q_3}$. The ${\cal S}{\cal
    S}^*$ pair invariant masses are in red, the $X_{\bar \q_{1,2}
    \q_3}$ HC multipion invariant masses are in blue, and the
  individual invariant mass distributions of the ${\cal S}$ and ${\cal
    S}^*$ are similar and shown in green.}\label{fig:uQutQt}
\end{figure}

The ${\cal S}{\cal S}^*$ production cross section is $2.7$\,pb in the
narrow-width approximation. Taking into account the large ${\cal S}$
decay width ($\Gamma_{\cal S} \approx 0.44 M_{\cal S}$), we find that
the $pp\to {\cal S}{\cal S}^*\to u_i \bar u_j X_{\bar \q_i \q_j}$
cross section is reduced to $1.4$\,pb in ${\tt MadGraph}$. The
dominant contributions are $pp\to {\cal S}{\cal S}^*\to q_{1,2} \bar t
\,X_{\bar \q_{1,2} \,\q_3}$ and the CP conjugate modes, with a total
cross section of $0.66$\,pb, and $pp\to {\cal S}{\cal S}^*\to t \bar t
X_{\bar \q_3 \q_3}$ with a cross section of $0.57$\,pb. In
Figure~\ref{fig:uQutQt} we show the mass distributions for $pp\to
{\cal S}{\cal S}^*\to q_{1,2} \, \bar t\, X_{\bar \q_{1,2} \q_3}$; the
distributions for the other decay modes of the ${\cal S}$ are very
similar. One can see that the bulk of the $X_{\bar \q_{1,2} \q_3}$
system has invariant masses that lie above $\Lambda_\chi\sim {\mathcal
  O}(250)$\,GeV, and also well above the threshold for multi-pion
production.  There is enough energy available to produce $tjK$,
$tjK+\pi$, $tjK+2\pi$, etc, multi-pion final states. The cross section
for producing one, two, three, or more HC pions depends on the details
of the HC dynamics, and thus on the hadronization model.  One could
contemplate rescaling the hadronization models used for QCD to the HC
scale to obtain a more quantitative description. However, for our
purposes a qualitative picture suffices.

If we were to model the hadronization of the $\bar \q_{1,2}\q_3$ pair
with a string model, the extra pions would be created from string
breaking. Since there is sufficient energy, the penalty for creating
an extra pion is small.  As we saw, pair production of ${\cal S}{\cal
  S}^*$ would result in $tj+n$ fat jets (the HC pions and kaons) or
$t\bar t+n$ fat jets final states. Here $n$ can lie anywhere from 1 to
${\mathcal O}(10)$. The ${\cal S}{\cal S}^*$ pair can thus be searched
for in multijet final states. Both the CMS and ATLAS collaborations
have recently made significant progress in multijet searches
\cite{ATLAS:2012dp, Chatrchyan:2013gia,
  ATLAS-CONF-2013-091}. Particularly relevant in this respect is the
ATLAS search \cite{ATLAS-CONF-2013-091}, which was interpreted in
terms of gluino production with R-parity violating decays that result
in either 6-jet or 10-jet final states (in the 6-jet search an extra
initial-state radiation jet was required in order to optimize the
sensitivity). Most importantly, the search strategy did not require
the jets to form resonances of a particular mass, and can thus be used
to place bounds on the production of wide resonances, such as ${\cal
  S}{\cal S}^*$. For a $520$\,GeV gluino that decays to $tjj$ the
ATLAS bound is $\sigma(pp\to \tilde g \tilde g \to \bar t t
+4j)<0.9$\,pb. This bound lies above the cross section for
$\sigma(pp\to {\cal S}{\cal S}^*\to t\bar t+n\pi)$ in our model.

The ATLAS collaboration has also performed a search in which the final
state contains $t\bar t+2b$ jets and a number of light jets. This
final state arises in our model from $pp\to {\cal S}{\cal S}^*\to
t\bar t \,K\bar K$ plus any number of other pNGBs (a much smaller
contribution could come from the $\phi$ resonance in place of $K\bar
K$). Here, the hadronization of HC quarks results in a $K\bar K$
pair. The kaons then decay to off-shell $t^*$, so that $K\to t^*j\to b
\, 3j$. We thus have $pp\to {\cal S}{\cal S}^*\to t\bar t \, 2b \,
6j$. The ATLAS search \cite{ATLAS-CONF-2013-091}, with a gluino
decaying to five quarks through an intermediate neutralino, gives an
upper bound of about $1.5$\,pb, well above our production cross
section of $0.6$\,pb.

\section{Electroweak precision tests, Higgs couplings\label{sec:EWK}}%
\begin{figure}
  \centering
  \includegraphics[width=\columnwidth]{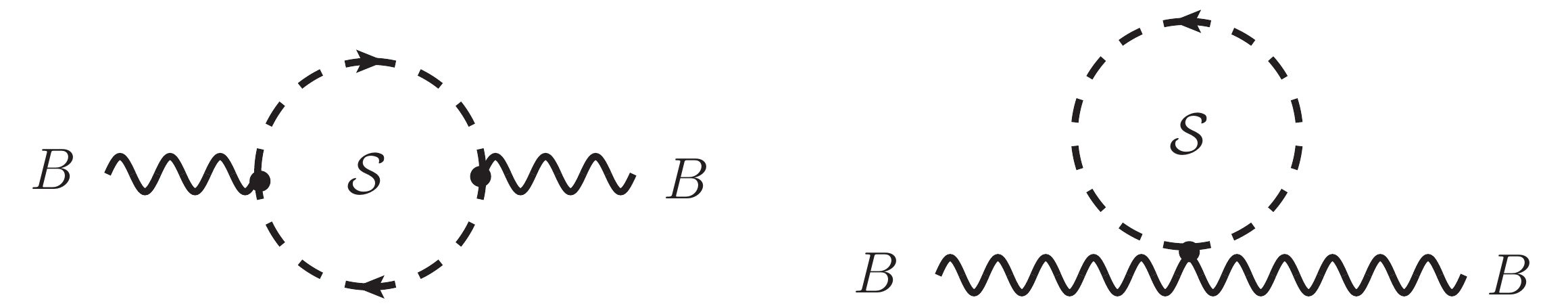}  
  \caption{The two contributions of the scalar $\mathcal S$ to the $T$
    parameter cancel. }
  \label{fig:T}
\end{figure}

In this section we discuss the implications of electroweak-precision
measurements for our model.  For the contributions to the electroweak
oblique parameters $S$ and $T$ due to new HC states we rely on the
operator product expansion (OPE) and quark hadron duality in order to
estimate these contributions.  Thus, the electroweak corrections are
given by the diagrams in Figure~\ref{fig:T} to good approximation. The
corrections are suppressed by powers of $\Lambda_{\rm HC} /m_{\cal
  S}$.  The scalar $\mathcal S$ is an ${\rm SU}(2)$ singlet and thus
does not contribute to the $S$ parameter. The contribution to the $T$
parameter, on the other hand, vanishes due to a cancellation between
the two diagrams shown in Figure~\ref{fig:T} (see also the discussion
of models with Higgs singlets in Ref.~\cite{Grimus:2007if}).  The HC
quarks are hypercharge singlets and thus do not contribute at this
order.

In Ref.~\cite{Gresham:2012wc} atomic-parity violation was advocated
as a strong constraint on t-channel explanations of the
forward-backward asymmetry. 
Below the electroweak scale atomic-parity violation can be described
by an effective electron-quark interaction of the form
\begin{equation}
{\mathcal L} = \frac{G_F}{\sqrt{2}} \sum_{q=u,d} \big( C_{1q} \bar e
\gamma^\mu \gamma_5 e \bar q \gamma_\mu q + C_{2q} \bar e \gamma^\mu e
\bar q \gamma_\mu \gamma_5 q \big)\, ,
\end{equation}
where the second term is suppressed by the small electron weak charge
and neglected in the following. We define the $Z$--light-quark
couplings as in Ref.~\cite{Gresham:2012wc} by
\begin{equation}
{\mathcal L} = - \frac{e}{s_w c_w} Z^\mu \big( a_R^\text{NP}(q)
\bar q_R \gamma_\mu q_R + a_L^\text{NP}(q) \bar q_L \gamma_\mu q_L
\big)\, . 
\end{equation}
In terms of the effective electron-quark Wilson coefficients we have
$C_{1q}^\text{NP} = a_L^\text{NP}(q) + a_R^\text{NP}(q)$.

\begin{figure}
  \centering
  \includegraphics[width=4cm]{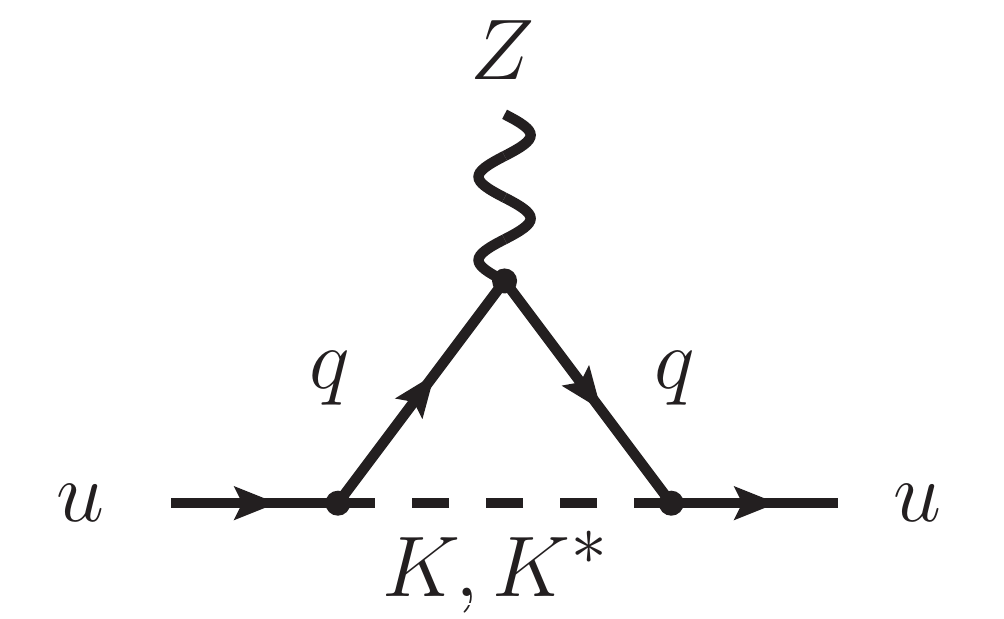}~~~~ 
  \includegraphics[width=4cm]{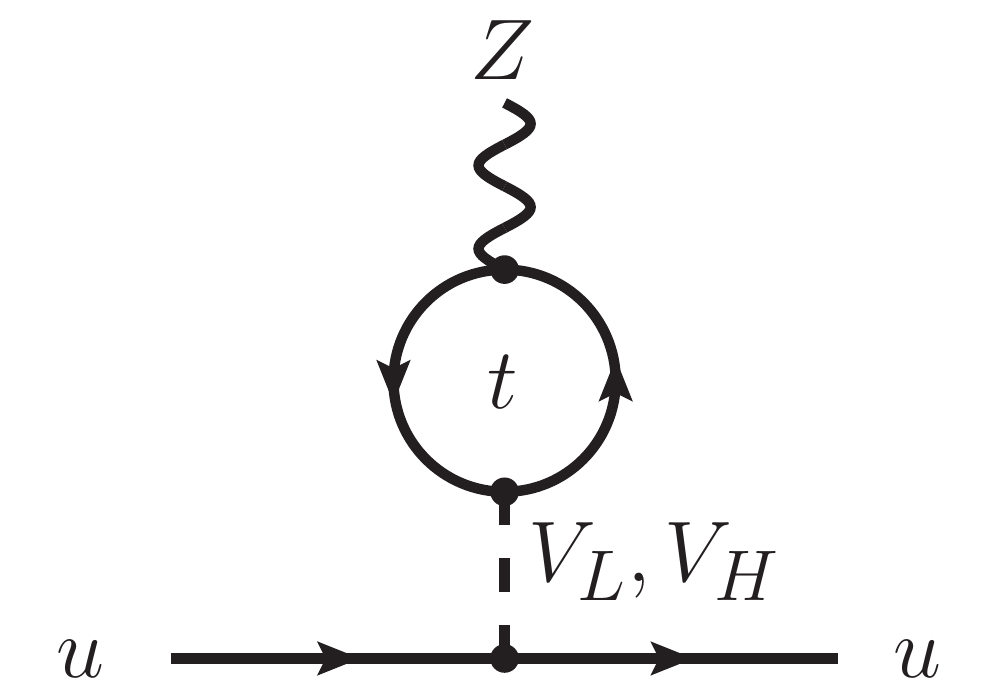}  
  \caption{Vector and pseudoscalar meson contribution to atomic-parity violation.}
  \label{fig:apvvec}
\end{figure}

To estimate the effect of the $K^*$ resonances on atomic-parity
violation, we compute the matching corrections to $a_R^{K^*}(u)$. To
this end we evaluate the diagrams with the exchange of a massive
vector and a top quark (cf. Figure~\ref{fig:apvvec}). The finite part
of the contribution of the $K^*$--top-quark loops, including field
renormalization, is
\begin{equation}
\begin{split}
a_R^{K^*}(u) & = 
 \frac{g_\rho^2 \sin^2 \theta_{R,1} \sin^2 \theta_{R,3}}{32 \pi^2} 
\\ & \quad \times \Big[
 \frac{x^2-7x}{8(x-1)} + \frac{3x\log x}{4(x-1)^2} -\frac{x}{4}
  \log\frac{\mu^2}{m_t^2} \Big] \,, 
\end{split}
\end{equation}
where $x\equiv m_t^2/m_{K^*}^2$. Note that the $K^*$ contribution is
divergent because of our use of a non-gauge vector propagator. The
divergent contribution vanishes in the limit $x \to 0$. We estimate
the size of the effect by varying the renormalization scale, $\mu$, in
the range $[M_{K^*}/2, 2M_{K^*}]$.  The same expression also applies
for $K_1$ exchange with $x\to m_t^2/m_{K_1}^2$ and $g_\rho\to
g_{a_1}$.

The effect of HC $K$ exchange is similarly estimated by evaluating the
diagram in Figure~\ref{fig:apvvec} with $K$ in the loop. The finite
part of the contribution of the $K$--top-quark loops, including field
renormalization, is given by
\begin{equation}
\begin{split}
 a_R^{K}(u) & =
 \Big( \frac{g_A^\text{HC}}{f_\pi^\text{HC}} \Big)^2 \frac{\sin^2
   \theta_{R,1} \sin^2 \theta_{R,3}}{32 \pi^2} M_K^2 \\ & 
  \quad \times \Big[ \frac{x+x^2}{8(1-x)} + \frac{x\log x}{4(x-1)^2} +
    \frac{x}{4} \log\frac{\mu^2}{m_t^2} \Big] \, ,
\end{split}
\end{equation}
where $x\equiv m_t^2/M_{K}^2$. Note that this contribution is
divergent because of the dimension-five couplings in
Eq.~\eqref{eq:pionL}. As in the case of the $K^*$ above, the divergent
contribution vanishes in the limit $x \to 0$, and the size of the
effect is estimated by varying the renormalization scale in the range
$[M_{K}/2, 2M_{K}]$. All contributions of the $K^* $ and $K$ loops
proportional to the weak mixing angle vanish after renormalizing the
external fermion fields.

The contribution of the top-quark loop diagram in
 Figure~\ref{fig:apvvec} (right) is given by 
\begin{equation}
\begin{split}
\label{eq:omegaphidiag}
a_R^{V_{H,L}}(u) & = \mp
 \frac{g_\rho^2 \sin^2 \theta_{R,1} \sin^2 \theta_{R,3}}{32 \sqrt{2} 
   \pi^2} \sin \theta_{V_{H,L}}^\text{id} \cos \theta_{V_{H,L}}^\text{id} \\ 
 & \quad \times N_c
 \Big[ \frac{m_t^2}{m_{V_{H,L}}^2}
  \log\frac{\mu^2}{m_t^2} \Big] \,.
\end{split}
\end{equation}
This is substantially suppressed by the small deviation from the ideal
$\omega-\phi$ mixing, $\sin\theta_{V}^{\rm id} = 0.028$ defined in
Eq.~\eqref{Vmixing} (see also Appendix~\ref{app:fit:to:spectra}).  The
analogous contribution from axial $A_L - A_H$ exchange is obtained by
replacing $\theta_{V_{H,L}}^\text{id}\to \theta_{V,A}^\text{id}$,
$g_\rho\to g_{a_1}$, $m_{V_{H,L}}\to m_{V,A}$. As above, the scale
$\mu$ in Eq.~\eqref{eq:omegaphidiag} and the axial-vector analog is
varied in the range of half to twice the resonance mass.

\begin{figure}
  \centering \includegraphics[width=4cm]{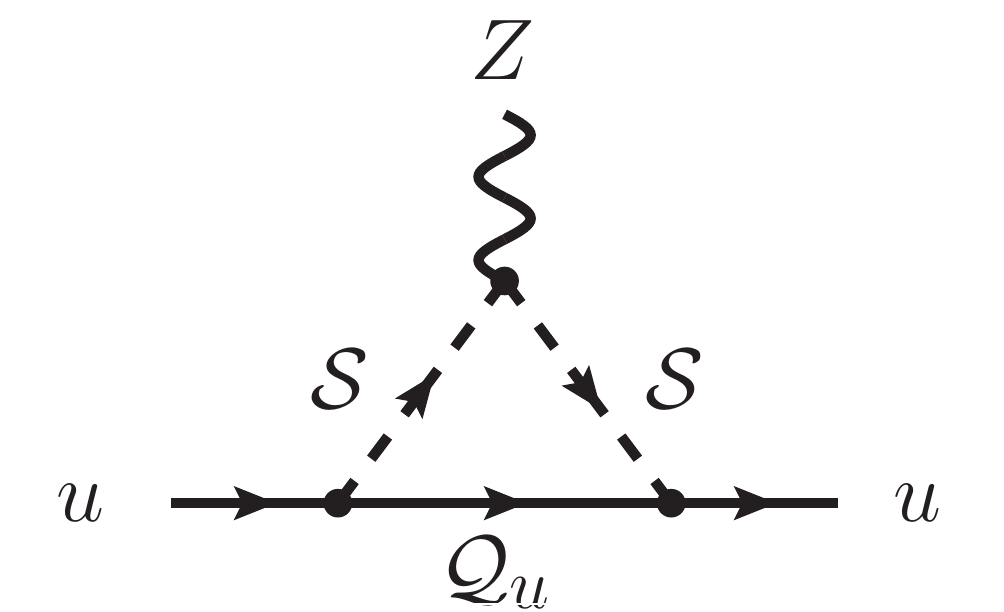}
  \caption{Contribution of the heavy scalar ${\cal S}$ to atomic-parity violation.}
  \label{fig:apvscal}
\end{figure}

Finally, we estimate the effect of the would-be composite quarks on
atomic-parity violation. As discussed in
Section~\ref{subsec:compquarks}, they are closely analogous to a
heavy-light meson. In our case the role of the heavy quark is played
by the heavy scalar. We thus evaluate the corresponding loops in the
UV theory, as shown in Figure~\ref{fig:apvscal}. The $Z$ coupling to
the scalar is given in terms of the covariant derivative
\begin{equation}
D_\mu = \partial_\mu + \frac{2}{3} \frac{ies_w}{c_w} Z_\mu  ,
\end{equation}
in the kinetic term 
\begin{equation}
{\cal L}_\text{kin} = \big( D_\mu {\cal S} \big)^\dagger D^\mu {\cal
  S} \, .
\end{equation}
We find that the contribution of the renormalized diagram
vanishes. 

We can obtain a bound on the size of $a_R^\text{NP}(u)$ from the
measurement of the nuclear weak charge of cesium
($^{133}\text{Cs}$). The contribution of $a_R^\text{NP}(u)$ to the
nuclear weak charge is given by
\begin{equation}
\Delta Q_W = -2 (2 Z + N) a_R^\text{NP}(u) \, ,
\end{equation}
where $Z$ and $N$ are the number of protons and neutrons in the
nucleus. From the difference of the experimental value,
$Q_W^\text{exp} = -72.58(43)$~\cite{Dzuba:2012kx}, and the central
value of the SM prediction $Q_W^\text{SM} = -73.23$ (calculated with
input from Ref.~\cite{Beringer:1900zz}),
\begin{equation}
\Delta Q_W (\text{Cs}) \equiv Q_W^\text{exp} (\text{Cs}) -
Q_W^\text{SM} (\text{Cs}) \subset [0.22, 1.08] ,
\end{equation}
we obtain the allowed 1-sigma region $a_R^\text{NP}(u) \subset [-0.28,
  -0.06] \%$. This should be compared with the HC resonance
contributions listed in Table~\ref{tab:bmres}.  One sees that the HC
effects are well within the errors on $a_R^{\rm NP}$.

The HC interactions modify the Higgs production cross sections and
decays branching ratios. However, we find these modifications to be
small in size. The effects of the quartic Higgs coupling to scalar
${\cal S}$ are suppressed by its large mass, $m_{\cal S}$, and are
negligible. Modifications of the Higgs couplings to the $W$ and $Z$
arise at loop level and are irrelevant.  In principle the partial
compositeness of the RH top quark could lead to appreciable
modifications in $\bar tth$ production, $gg\to h$ fusion, and the
$h\to\gamma\gamma$ decay channel, via the RH and LH mixings in
Eqs.~\eqref{eq:mixinganglesRH}, \eqref{eq:mixinganglesLH}.  The $\bar
t th$ production cross section is given by
\begin{equation}
\frac{\sigma_{\bar t th}^{\rm NP}}{\sigma_{\bar t th}^{\rm
    SM}}=\Big(\frac{y_t^{\rm NP}}{y_t^{\rm
    SM}}\Big)^2\cos\theta_{R3}^2\cos\theta_{L3}^2=1+{\mathcal
  O}(1/M^4)\,, 
\end{equation}
where $y_t^{\rm NP}\equiv m_{u_3}/v$ is the top-quark Yukawa coupling
in the interaction basis, which differs from the SM relation $y_t^{\rm
  SM}=m_t^{\rm phys}/v$. Since the physical top-quark mass is given by
$m_t^{\rm phys}\simeq m_{u_3} \cos\theta_{R3}\cos\theta_{L3}$ the net
change in the $t\bar th$ production cross section is small.
Numerically, it is an ${\mathcal O}(1\%)$ effect.

In the limit of a heavy top, where the Higgs low-energy theorem
applies, the contributions of the top and the would-be composite top
quark $u'_3$ running in the loop completely cancel in the $gg\to h$
and $h\to \gamma\gamma$ amplitudes.  The net modifications of the
gluon-fusion cross section and the $h\to \gamma\gamma$ branching ratio
therefore lie well below a percent.

\begin{table}
\centering
\begin{tabular}{ll}
\hline\hline
		& 1-$\sigma$ range [\%] \\
		\hline
$a_R^{K}(u)$ 	&$[-0.215, 0.010]$\\
$a_R^{K^*}(u)$ 	&$[-0.037, -0.005]$\\
$a_R^{K_1}(u)$ 	&$[-0.054, -0.023]$\\
$a_R^{V_L}(u)$ 	&$[-0.0008, 0.0009]$\\
$a_R^{V_H}(u)$ 	&$[-0.0002, 0.0007]$\\
		\hline\hline
\end{tabular}
\caption{Range of the effective $a_R$ coupling to $Z$ in \%, induced
  by the HC resonance contributions discussed in the main text. }
\label{tab:bmres}
\end{table}

\section{The signals of stealth strong dynamics\label{sec:signals}}%
Our strong interaction model for enhanced $t\bar t$ asymmetries makes
several predictions that are not tied to the exact numerical values of
the UV parameters and are thus quite robust. It predicts the existence
of a tower of resonances that couples strongly to the right-handed top
$t_R$: $K, K^*, K_1,...$ There is a flavor octet of pNGBs: $\pi, K,
\eta$ (plus the $\eta'$), with the lightest state decaying to two
jets. The latter is most likely a triplet of pions, with a mass of
$\sim 50$\,GeV. Finally, the minimal form of the model also predicts
the existence of a stable, electrically neutral ``HC baryon'' with a
mass of $\sim 250$\,GeV, which may be searched for in direct dark
matter (DM) detection experiments.

\begin{figure}[]
\begin{center}
\hspace*{-1em}\includegraphics[]{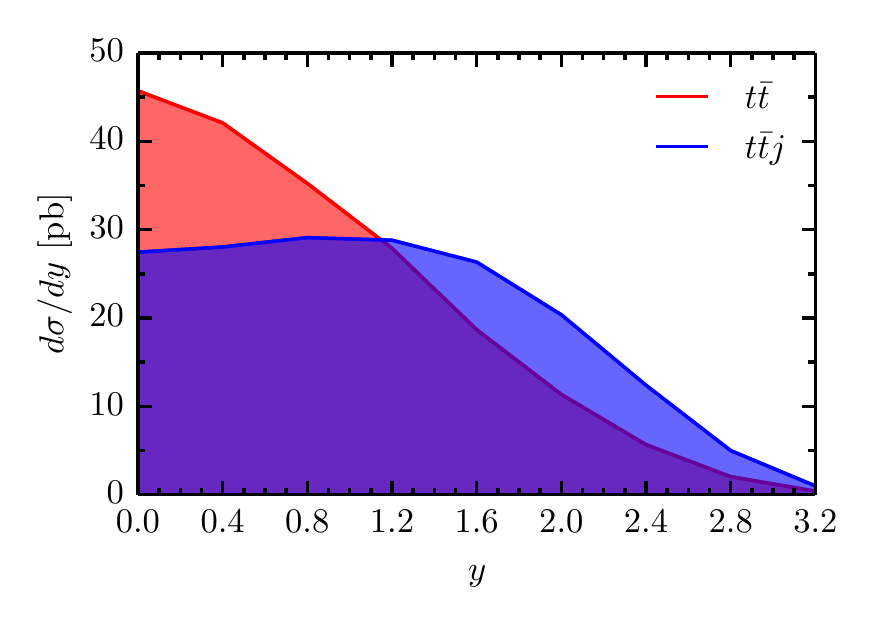}
\end{center}
\caption{The rapidity distribution of the anti-top quark from $pp \to
  t \bar t$ (red) and $pp \to t K^* \to t \bar t j$ (blue), for LHC at 13 TeV. The
  $pp \to t \bar t j$ differential cross section has been rescaled by
  a factor $20$.  
}\label{fig:ytbar}
\end{figure}

The HC resonances $K^*, K_1, ...$ decay to $t+j$ final states and are
already being searched for, as discussed in
Section~\ref{new-res}. Their production cross section will increase
roughly $4$-fold in going from the 8-TeV LHC to the 13-TeV LHC, see
Table~\ref{tab:qprime}. This should be compared with the corresponding
$\sim 5$-fold increase in the $t\bar t$ cross section. The challenge
will be to search for $t+j$ resonances given the larger hadronic
activity in 13-TeV events. One could explore the fact that, at 13~TeV,
the anti-top quarks in $pp\to t + K^* \to t \bar{t} j$ will in general
be produced at larger rapidities than the anti-top quarks in $t\bar t$
events (see Figure~\ref{fig:ytbar}).  The usefulness of this charge
asymmetry at the LHC has been discussed in Ref.~\cite{Knapen:2011hu}
for the case of associated $W' t $ production.

Next, we discuss searches for HC $K$ mesons. A challenge here is that
the $K$ decays to an off-shell top, $K\to t^*j$. Ideally, the present
experimental searches would be optimized to search not just for $t+j$
resonances but also for $t^*+j$ resonances. Gains are potentially
possible, if one allows for softer leptons from the semileptonic
decays of the $t^*$.

The discovery of a light pion decaying to two jets would be a
particularly striking signal of stealth strong dynamics. The challenge
in searching for HC pions is that they are most copiously produced in
decays of higher resonances, which typically results in high
multiplicity final states. An exception is the HC $\rho$ resonance,
which almost exclusively decays through $\rho\to \pi\pi$. The
s-channel production $pp\to \rho\to\pi\pi$, with $\pi\to 2j$ is
effectively already searched for in paired dijet events, as discussed
in Section~\ref{sed:dijets}. However, in our model both $\rho$ and
$\pi$ are very light, thus present searches are not sensitive to them.
It is unlikely that the sensitivity to the low-mass region can be
improved in this type of search at the LHC with increased collision
energies.  Potentially more promising may be the pair production of
$\rho$ resonances, $pp\to \rho \rho\to 4\pi$, in which each pion
decays to two jets. Since the $\rho$'s would come primarily from the
partonic $u\bar u\to \rho\rho$ process, with the $u$ quark in the
t-channel, they would have large rapidities.

The cross sections for the process $p p \to \rho \rho$ are $0.38$\,pb
at LHC8 and $1.38$\,pb at LHC13, respectively.  A promising search
strategy would thus be to optimize a search for forward pair
production of resonances, resulting in two fat jets (potentially
further resolved into two jets each using jet substructure
techniques).

Finally, our model in its minimal form of Eq.~\eqref{eq:LagHC}
contains a stable neutral HC baryon $B_\chi$ formed from three of the
light HC quarks.  The Lagrangian is invariant under a global ${\rm
  U}(1)_{{\rm \cal B}_{\rm HC}} $ HC baryon-number symmetry, under
which the HC quarks $\q_i$ have charge ${\rm \cal B}_{\rm HC} = 1/3$,
${\cal S}$ has charge ${\rm \cal B}_{\rm HC} = -1/3$, while all gauge
and SM-matter fields are neutral.  Alternatively, one can consider the
$Z_2^{\rm HC}$ subgroup, under which the HC quarks and ${\cal S}$ are
odd, and the SM matter is even. The lightest HC baryon state $B_\chi$
(${\rm \cal B}_{\rm HC} =1$, or $Z_2^{\rm HC}$ odd) is therefore
stable.  We estimate its mass by rescaling from QCD, yielding
$m_{B_\chi } \sim m_p \,f_\pi^{\rm HC} / f_\pi^{\rm HC} \sim
220$\,GeV, not including the HC quark-mass contributions.

Small breaking of the flavor ${\rm U}(2)_{U_R}$ symmetry should be
accompanied by small mass splittings between the light HC quarks, with
$m_{\q_2} > m_{\q_1}$ or $m_{\q_2} < m_{\q_1}$.  We therefore consider
two possible flavor structures for the lightest HC baryon, $B_\chi
[\q_1\q_1\q_2]$ or $B_\chi [\q_2\q_2\q_1]$, respectively.  In general,
we expect the two baryons in this ``isospin doublet'' to be nearly
degenerate in mass, since the ${\rm U}(2)_{U_R}$ symmetry must remain
approximately intact due to FCNC constraints.

If $B_\chi$ is a thermal relic, its relic abundance is set by its
annihilation cross section in the early universe. This is dominated by
$B_\chi \overline{B}_\chi \to $ multi-$\pi_{\rm HC}$ final states.  We
estimate this by scaling the QCD $p\bar p$ annihilation cross section
to the HC scale,
\begin{equation} 
\sigma_{B_\chi \overline B_\chi } \approx \sigma_{p \bar p} \,
(f_\pi^{\rm HC} /f_\pi )^2 \,.
\label{eq:crosssecscale}
\end{equation}
The $p\bar p$ annihilation cross sections measured by the LEAR
collaboration \cite{Bruckner:1989ew} vary from $\approx 200$\,mb to
$\approx 80$\,mb, for $\bar p$ beam momenta varying from $200$\,MeV to
$\approx 600$\,MeV, respectively.  Using Eq.~\eqref{eq:crosssecscale}
and converting the LEAR $\bar p$ beam momenta to center of mass values
of $v/c$, we obtain the range of $B_\chi$ annihilation cross sections
$\sigma_{B_\chi \overline B_\chi } \approx 0.01$\,mb ($v/c \approx
0.1$) to $\sigma_{B_\chi \overline B_\chi } \approx 0.004$\,mb ($v/c
\approx 0.3$).  Note that the non-relativistic values of $v/c$ are in
the range relevant for estimating thermal relic DM abundances.  In
particular, we find $\langle \sigma_{B_\chi \overline B_\chi } \,v
\rangle \sim 3 \cdot 10^{-20}~{\rm cm^3 /s }$.  This is $10^6 \times
\langle \sigma v\rangle_{\rm th. rel.}$, where $\langle \sigma
v\rangle_{\rm th. rel.}= 3\cdot 10^{-26} \text{cm}^3/s$ would be the
annihilation cross section required for $B_\chi$ to explain the
observed DM abundance.

Our estimate for $\langle \sigma_{B_\chi \overline B_\chi } \,v
\rangle$ implies that the $B_\chi$ would be a very subleading DM
component, with $\Omega_{B_\chi}\sim {\mathcal
  O}(10^{-6})\times\Omega_{\rm DM}$. Nevertheless, DM direct detection
searches could be sensitive to it, because the $B_\chi$-nucleon
scattering cross section is large.  The latter is dominated by
exchange of the vector mesons $\rho_{HC}, V_L ( \omega_{\rm HC}) $.
For simplicity, we only consider the vector couplings $ g_V \overline
B_\chi \gamma_\mu B_\chi V^\mu $, and ignore the higher-dimension
tensor couplings.  For the coupling strengths, we use the
corresponding QCD sum-rule values \cite{Erkol:2006sa}: $g_\rho = 2.3$
and $g_\omega = 6.9$.

For $B_\chi  [\q_1\q_1\q_2]$, 
its cross section with protons is given by 
\begin{equation}
  \sigma (B_\chi \,p \to B_\chi \,p )  =
  (\kappa_R^\rho)^2{(g_\rho +  g_{\omega} )^2 \over m_{\rho_{\rm
        HC}}^4 }  {m_p^2  \over 16 \pi} \,, 
\label{eq:Bchipxsec}
\end{equation}
where we have ignored the small splitting between the $\rho_{\rm HC}$
and $V_L$ masses, and the couplings of the $\rho_{\rm HC} $ and $V_L$
to the RH up quark are equal and given by $\kappa^\rho_{R} = 0.117$,
see Table~\ref{tab:quark-couplings}.  For $B_\chi [\q_2 \q_2 \q_1]$,
its cross section with protons is obtained by substituting $g_\rho \to
- g_\rho$ in Eq.~\eqref{eq:Bchipxsec}. In both cases the cross
sections with neutrons are a factor of four smaller.  Thus, for our
benchmark we estimate that the $B_\chi$--nucleon cross section would
be $\sigma_{\rm scatt} \sim 10^{-38}~{\rm cm}^2$ for $B_\chi [\q_2
  \q_2 \q_1]$, and $\sigma_{\rm scatt} \sim 5 \cdot 10^{-38}~{\rm
  cm}^2$ for $B_\chi [\q_1 \q_1 \q_2]$.  The LUX bound on the
DM-nucleon scattering cross section excludes
$(\Omega_{B_\chi}/\Omega_{\rm DM})\times \sigma_{\rm scatt}\lesssim
2.5 \cdot 10^{-45}\, \text{cm}^2$~\cite{Akerib:2013tjd} for
$m_{B_\chi}\sim 220$\,GeV.  With our estimate for
$\Omega_{B_\chi}/\Omega_{\rm DM}$, the scattering of relic $B_\chi
[\q_2 \q_2 \q_1] $ on nuclei would seem to be at the very limit of the
allowed range, whereas for $B_\chi [\q_1 \q_1 \q_2] $ it is an order
of magnitude too large.  The uncertainties in our estimates of the
$B_\chi$ relic density and cross sections with nucleons are large and
warrant a more detailed analysis to see whether or not stealth strong
dynamics could be discovered in direct DM searches.

Finally, we point out that the HC baryon number (or discrete $Z_2^{\rm
  HC}$) symmetry is accidental and therefore may only be
approximate. For instance, it could be broken at some higher scale in
extensions of our model with a larger field content. In that case all
of the HC baryons could be unstable and decay in the early
universe. For example, if the SM is extended by a light right-handed
neutrino, $\nu_R$, the breaking can exhibit itself through a
dimension-seven operator $({\cal S}^\dagger\q_1\q_2) u_R \nu_R$ which
also breaks the ${\rm U}(2)_{U_R}$ symmetry, where the ${\rm
  SU}(N)_{\rm HC}$ indices in the bracket are contracted with the
antisymmetric Levi-Civita tensor.

\section{Conclusions\label{sec:Conclusions}}%
We have provided an explicit strong interaction model that can produce
large enhancements of the $t\bar t$ asymmetries at the Tevatron, while
not being excluded by other direct or indirect searches for new
physics.  The model minimally extends the SM field content by a (SM
gauge singlet) flavor triplet of vectorlike fermions $\q_i$ charged
under a new strong gauge group, and by a scalar ${\cal S}$ charged
under QCD, hypercharge and the new strong gauge group.  We choose
${\rm SU}(3)_{\rm HC}$ ``hypercolor'' for the strong gauge group in
order to directly translate the results of QCD strong dynamics, thus
reducing the uncertainties in our predictions.  An approximate ${\rm
  U}(2)$ flavor symmetry is imposed on the fundamental interactions
between the hypercolor and SM sectors -- the Yukawa-type couplings of
the ${\cal S}$ to the RH up quarks and $\q_i$.  The $\q_i$ masses are
also ${\rm U}(2)$ invariant, and satisfy a hierarchy analogous to
$m_{u,d} \ll m_s \sim f_\pi$ in QCD.  The ${\cal S}$ is heavier, and
is thus analogous to a heavy flavor quark in QCD with mass $m \gg 4
\pi f_\pi$.  The flavor symmetry insures that new physics
contributions to flavor-changing neutral currents, e.g.,
$D^0$--$\overline{D^0}$ mixing or same-sign top pair production, are
negligible or absent.  The fact that the HC sector is neutral with
respect to the ${\rm SU}(2)_L $ weak interaction allows the model to
easily evade precision electroweak tests, and also ensures that any
modifications of the Higgs couplings are very small.

The scale of the new strong dynamics is quite low. The mass of the
lightest pseudo Nambu-Goldstone boson, $\pi_{\rm HC},$ is merely
$60$\,GeV, while most of the remaining resonances are in the range of
$\approx 150$\,GeV to $\approx 300$\,GeV.  Despite a plethora of new
resonances, the model is, surprisingly, not yet excluded by new
physics searches. The reason is two-fold: {\it (i)} these resonances
are bound states of the $\q_i$ and are thus QCD color neutral, {\it
  (ii)} the heavier hypercolor scalar ${\cal S}$ rapidly decays into
high-multiplicity final states, before it can form QCD colored bound
states.  In particular, pair production of ${\cal S}{\cal S}^*$ would
result in $tj+n$ fat jets or $t\bar t+n$ fat jets final states, where
a ``fat jet'' is associated with a $\pi_{\rm HC}\to jj$ or $K_{\rm HC}
\to t^* j$ decay.

There already has been considerable progress at the LHC in searches
for new physics involving final states with large jet multiplicities.
However, searches in relatively low-multiplicity final states,
including fat jets, are probably the most promising for discovery of
hypercolor resonances.  An example is pair production of $\rho_{\rm
  HC}$ resonances, with each $\rho_{\rm HC}$ decaying to a pair of HC
pions, resulting in the chain $pp\to \rho_{\rm HC} \,\rho_{\rm HC} \to
4\,\pi_{\rm HC} \to 4$ fat jets.  The rather strong coupling of the
hypercolor sector to the top quarks leads to observable $t\bar t$
charge asymmetries in associated hypercolor resonance production at
the LHC.  This feature could be useful in discriminating signal from
background in $t$+ jet resonance searches.  For example, the $\bar t$
in the process $pp \to t K_{\rm HC}^* \to t \bar t j $ is produced
with relatively large rapidities compared to the top, and compared to
the anti-top in $pp \to t\bar t$.  This feature could also be useful
in virtual $t^* + j$ resonance searches associated with the process
$pp \to t K_{\rm HC} \to t \bar t^* j $.

We have used known non-perturbative aspects of QCD, as well as
familiar approximations like QCD sum rules, vector-meson dominance,
and a naive quark model, combined with simple scaling arguments to
obtain reasonable estimates of the resonance masses and interaction
strengths in our QCD-like hypercolor model.  An interesting
application concerns the relic abundance of the lightest HC baryon,
$B_\chi$. If the accidental HC baryon number symmetry of our
Lagrangian is left unbroken by higher-dimensional operators involving
additional light non-SM fields, then the $B_\chi$ is stable.  Scaling
the measured low-energy QCD $p\bar p$ annihilation cross sections will
imply a $B_\chi$ relic abundance that is only ${\mathcal O}(10^{-6})$
as large as the observed dark matter abundance.  Nevertheless, the
$B_\chi$--nucleon cross section is large enough to be close to
saturating the present direct dark matter detection bounds, and may
yield an observable signal in the next generation of the dark matter
direct detection experiments or rule out a stable $B_\chi$.

An interesting open question is also the UV structure of our model.
The one-loop running of the Yukawa couplings $h_1 , h_3$ and the HC
gauge coupling $g_{\rm HC}$ gives a Landau pole at $\mu = O({\rm
  few~TeV})$, whereas two-loop RGEs give an approximate UV fixed point
for $h_3 $ of non-perturbative strength, $h_3^* \approx 8$, that is
realized at $\mu = O(10$\,TeV), with $h_1$ and $g_{\rm HC}$
asymptotically free.  To settle the question which of the two
possibilities is realized would require nonperturbative methods, and
is beyond the scope of the current work, but could be an interesting
future research direction.

In summary, the presented model can lead to large $t\bar t$ asymmetry
at the Tevatron, evades present experimental bounds, but can be
searched for with improved strategies at the LHC and in direct dark
matter detection experiments.

{\bf Acknowledgements:} J.B. and E.S. would like thank Antonio Pich
for useful discussions. We would like to thank the Weizmann Institute
Phenomenology group, and the CERN theory group for their warm
hospitality during various stages of this work.  A.\,K. and J.B. are
supported by DOE grant FG02-84-ER40153.  J.\,Z. and J.B. are supported
by the U.S. National Science Foundation under CAREER Grant
PHY-1151392.  We also thank the Aspen Center for Physics, supported by
the NSF Grant \#1066293, and the KITP, supported in part by the
National Science Foundation under Grant No. NSF PHY11-25915, for their
warm hospitality. A.K. and J.Z. are grateful to the Mainz Institute
for Theoretical Physics (MITP) for its hospitality and its partial
support during the completion of this work. E.S. would like to thank
the Department of Physics at University of Cincinnati for its warm
hospitality during the initial stages of this work.

\appendix
\pdfbookmark[1]{Appendix}{appendix}
\section{Resonance mass spectra\label{app:fit:to:spectra}}%
The masses of the HC mesons are obtained from the measured spectrum of
QCD mesons by appropriate rescalings.  To good approximation, the QCD
$\rho$ mass, $m_\rho^{\rm QCD}$, corresponds to the massless limit of
the HC $\rho$ meson, $M_\chi=\lim_{m_{\q_i}\to 0 }\,m_\rho^{\rm HC}$.
In obtaining the spectra we also need to allow for variatons of the HC
quark masses. We do this by employing ChPT for the pseudoscalar mesons
and a naive quark model for the vector and axial-vector resonances.

\subsection{Pseudoscalar mesons}%
The compositions of the HC pion and kaon in terms of the HC quarks are
\begin{equation}
\begin{split}
|\pi^{1(2)} \rangle &= |\bar \q_{1(2)} \q_{2(1)} \rangle\,,\\
|\pi^{3} \rangle&={1\over \sqrt{2}} (|\bar \q_1 \q_1\rangle -|\bar \q_2 \q_2 \rangle)\,,\\
|K^{1(2)} \rangle &= |\bar \q_{1(2)}  \q_3 \rangle \,, \qquad
|\bar K^{1(2)} \rangle =  |\bar \q_3  \q_{1(2)} \rangle\,, 
\end{split}
\end{equation}
while the octet and the singlet pseudoscalars are given by
\begin{equation}
\begin{split}
|\eta_8 \rangle &= {1\over \sqrt{6}} (|\bar \q_1 \q_1\rangle + |\bar \q_2 \q_2 \rangle - 2  |\bar \q_3 \q_3 \rangle )\,,\\
|\eta_0 \rangle &= {1\over \sqrt{3}} (|\bar \q_1 \q_1\rangle + |\bar \q_2 \q_2 \rangle +  |\bar \q_3 \q_3 \rangle )\,.
\end{split}
\end{equation}
As explained in the main text, we neglect $\eta - \eta^\prime $
mixing, so that the $\eta_8=\eta$ and $\eta_0=\eta'$.

The quadratic terms in the chiral Lagrangian yield expressions for the
squared pseudoscalar masses which are linear in the quark masses
\begin{equation}
\begin{split}
  \left(m_{\pi^{1,2,3}}\right)^2 &= k_f \frac{M_\chi}{m^\text{QCD}_\rho}~2 m_{\q_1}\,, 				\\
  \left(m_{K^{1,2},\bar K^{1,2}}\right)^2 & = k_f \frac{M_\chi}{m^\text{QCD}_\rho}~(m_{\q_1} + m_{\q_3})\,, 		\\
  \left(m_{\eta}\right)^2        &= k_f \frac{M_\chi}{m^\text{QCD}_\rho}~\frac{2}{3}(m_{\q_1} + 2 m_{\q_3})\,. 
\end{split}
\end{equation}
We use $k_f= 2.765$ as obtained from the lattice QCD calculation in
Ref.~\cite{McNeile:2012qf}. The $\eta^\prime$ is assumed to be much
heavier than the other pseudoscalar mesons and is omitted from our
analysis as explained in Section~\ref{subsec:pseudoGB}.

\subsection{Vector Mesons}%
The vector meson masses and the mixing angle between the
flavor-singlet and octet states are calculated using the naive quark
model approach in Ref.~\cite{Cheng:2011fk}. The decompositions of the
$\rho$ and $K^*$ in terms of HC quark states are
\begin{equation}
\begin{split}
|\rho^{1(2)} \rangle &= |\bar \q_{1(2)} \q_{2(1)} \rangle\,,\\
~|\rho^{3} \rangle &={1\over \sqrt{2}} (|\bar \q_1 \q_1\rangle -|\bar \q_2 \q_2 \rangle)\,,\\
|K^{*1(2)} \rangle &= |\bar \q_{1(2)}  \q_3 \rangle \,, ~~
|\bar K^{*1(2)} \rangle =  |\bar \q_3  \q_{1(2)} \rangle\,.
\end{split}
\end{equation}
Their masses are given by
\begin{equation}
\begin{split}
  \left( m_{\rho^{1,2,3}}  \right)^2 &= \mu_V^\text{HC} \left( E^\text{HC}_{0,V} +2m_{\q_1} \right),\\
  \left( m_{K^{*1,2},\bar K^{*1,2}} \right)^2  &=\mu_V^\text{HC} \left( E^\text{HC}_{0,V} + m_{\q_3} +m_{\q_1} \right).
\end{split}
\end{equation}
Here $\mu_V^\text{HC}$ is an overall mass-scale parameter, $E_{0,V}$
describes the binding energy in the limit of massless HC quarks, while
the $m_{\q_i}$ are the HC quark masses.

The octet and singlet vector meson states are
\begin{equation}
\begin{split}
|V_8 \rangle &= {1\over \sqrt{6}} (|\bar \q_1 \q_1\rangle + |\bar \q_2 \q_2 \rangle - 2  |\bar \q_3 \q_3 \rangle )\,,\label{V8state}\\
|V_0 \rangle &= {1\over \sqrt{3}} (|\bar \q_1 \q_1\rangle + |\bar \q_2 \q_2 \rangle +  |\bar \q_3 \q_3 \rangle )\,.
\end{split}
\end{equation}
The mixing angle between the flavor-singlet and octet vector mesons is
obtained by diagonalising the corresponding mass matrix
\begin{widetext}
\begin{equation}
\begin{split}
  \begin{pmatrix}
    \left( m_{V_L} \right)^2	& 0				\\
    0					& \left( m_{V_H} \right)^2	\\
  \end{pmatrix}=
  R_V
  \begin{pmatrix}
    \mu_V^\text{HC} \left(E_{0V}^\text{HC}+\tfrac{2}{3}(2m_{\q_1}+m_{\q_3})+x_\text{an,V}^\text{HC}\right)	& 				
    -\tfrac{2}{3}\sqrt{2}\mu_V^\text{HC} (m_{\q_3}-m_{\q_1})					\\
    -\tfrac{2}{3}\sqrt{2}\mu_V^\text{HC} (m_{\q_3}-m_{\q_1})					&
    \mu_V^\text{HC} \left(E_{0V}^\text{HC}+\tfrac{2}{3}(m_{\q_1}+2m_{\q_3})\right)		\\
  \end{pmatrix}
  R_V^{-1},
\label{mVLH}
\end{split}
\end{equation}
\end{widetext}
where $V_{L,H}$ are the mass eigenstates. The mass matrix on the
r.h.s.~is given in the $V_0-V_8$ basis.  It contains an additional
parameter $x_{\text{an,V}}^{\rm HC}$ which takes into account the
annihilation of the flavor-singlet meson into gluonic intermediate
states. The matrix $R_V$ is chosen to diagonalize the mixing matrix
and thereby yields the mixing angle,
\begin{equation}\label{VLH}
\begin{pmatrix}
|V_L\rangle\\|V_H\rangle
\end{pmatrix}
=R_V\begin{pmatrix}
|V_{0}\rangle\\|V_8\rangle
\end{pmatrix}\,=
\begin{pmatrix}
\cos\theta_V&\sin\theta_V\\
-\sin\theta_V&\cos\theta_V
\end{pmatrix}
\begin{pmatrix}
|V_{0}\rangle\\|V_8\rangle 
\end{pmatrix}\, .
\end{equation}

We first fit the parameters of the ansatz in Eq.~\eqref{mVLH}, applied
to QCD. Our inputs are the measured QCD vector-meson masses, the
mixing angle $\theta_V^{\rm QCD}=38.7^\circ$, and the light-quark
masses $m_u=m_d=4$\,MeV, $m_s=100$\,MeV. They yield
\begin{equation}
\begin{split}
  \mu_V^\text{QCD} &= 2.21\,\text{GeV}\,, \quad E_{0,V}^\text{QCD} =
  260\,\text{MeV} \,, \\ 
  x_{\text{an},V}^\text{QCD} &= 15\,\text{MeV} \,,
\end{split}
\end{equation}
where we only quote the central values. This suffices since the errors
are much smaller than the errors we ascribe to the extrapolation to
the HC case. The HC parameters are obtained by rescaling in the usual
way
\begin{equation}
\{ \mu_V^\text{HC}, E_{0,V}^{\rm HX}, x_{\rm an, V}^{\rm HC} \}= 
\frac{M_\chi}{m^\text{QCD}_\rho} \{ \mu_V^\text{QCD}, E_{0,V}^{\rm QCD}, x_{\rm an, V}^{\rm QCD} \}\,,
\end{equation}
Note that $M_\chi=
\sqrt{\mu_V^\text{HC} E_{0,V}^\text{HC}}$. 

The mixing angle is close to ideal. The deviation from ideal mixing is
parametrized by the angle $\theta_V^{\rm id}$, see the definition in
Eq.~\eqref{Videal:def}. It is related to $\theta_V$ as
\begin{equation} 
\sin\theta_V^{\rm id}=-\frac{1}{\sqrt
  3}\cos\theta_V+\sqrt{\frac{2}{3}}\sin\theta_V.\label{thetaVid}
\end{equation} 
The singlet--octet mixing angle 
$\theta_V$ is directly related to $x_{\rm an, V}^{\rm
  HC}$ as ~\cite{Cheng:2011fk} 
\begin{equation} 
\tan 2 \theta_V =2
\frac{\sqrt{N_f-1}}{N_f-2 -\xi_V}\,, 
\end{equation}
with 
\begin{equation} 
\xi_V=\frac{N_f
  x_{\rm an, V}^{\rm HC}}{2 (m_{\q_3}-m_{\q_1})}
\label{xiV} 
\end{equation}
and $N_f=3$ in our HC model.  Note in particular that $x_{\rm an,
  V}^{\rm HC}=0$ corresponds to ideal mixing, $\theta_V^{\rm id}=0$.

\subsection{Axial vectors}%
The same naive quark model~\cite{Cheng:2011fk} can also be applied to
the axial ($^3P_1$) vector masses. The decompositions of the $a_1$ and
$K_1$ in terms of HC quark states are
\begin{equation}
\begin{split}
|a_1^{1(2)} \rangle &= |\bar \q_{1(2)} \q_{2(1)} \rangle\,,\\
~|a_1^{3} \rangle &={1\over \sqrt{2}} (|\bar \q_1 \q_1\rangle -|\bar \q_2 \q_2 \rangle)\,,\\
|K_1^{1(2)} \rangle &= |\bar \q_{1(2)}  \q_3 \rangle \,, \qquad
|\bar K_1^{1(2)} \rangle =  |\bar \q_3  \q_{1(2)} \rangle\,.
\end{split}
\end{equation}
Note that we have ignored the ${}^1P_1$ multiplet and the
corresponding $K_{1A}$-$K_{2A}$ mixing, as explained in the main text.
The naive quark-model parameters are now $\mu_A^{\rm HC}$,
$E_{0,A}^{\rm HC}$ and $x_{\rm an,A}^{\rm HC}$. The $a_1$ and $K_1$
masses are given by
\begin{equation}
\begin{split}
  \left( M^\text{HC}_{a_1^{1,2,3}}  \right)^2       &= \mu_A^\text{HC}\left( E^\text{HC}_{0,A} +2m_{\q_1} \right) ,		\\
  \left( M^\text{HC}_{K_1^{1,2}} \right)^2  &= \mu_A^\text{HC}\left( E^\text{HC}_{0,A} + m_{\q_3} +m_{\q_1} \right) .
\end{split}
\end{equation}

The description of the $A_0-A_8$ (singlet-octet) system is given by
substituting $V\to A$ in Eqs.~\eqref{V8state}-\eqref{VLH} and
\eqref{thetaVid}-\eqref{xiV}. The mass eigenstates are denoted by
$A_L$ and $A_H$.

The quark model parameters in QCD are obtained by fitting to the
$a_1$, $K_{1A}$, $f_1(1420)$, $f_1(1285)$ masses, and the mixing angle
$\theta_A^{\rm id}$.  For the mixing angle we inflate the errors,
taking $\theta_A^{\rm id}=(23.0\pm 23.0)^\circ$ (a recent LHCb
determination~\cite{Aaij:2013rja} obtains $\pm
(24^{+3.1+0.6}_{-2.6-0.8} )^\circ$ and recent lattice determinations
in Ref.~\cite{Dudek:2013yja} range from $\pm [25,36]^\circ$).  This
gives
\begin{equation}
\begin{split}
  \mu_A^\text{QCD} &= 2.16\,\text{GeV}\,, \quad E_{0,A}^\text{QCD} =
  700\,\text{MeV} \,, \\ 
  x_{\text{an},A}^\text{QCD} &= 64\,\text{MeV} \,,
\end{split}
\end{equation}
where again we only quote the central values.

\subsection{Would-be composite quarks ${u^\prime}$}%
The would-be composite quarks, $u_i'\sim [{\cal S}\q_i]$, can be
thought of as analogues of a heavy--light vector meson, as discussed
in Section~\ref{subsec:compquarks}. Carrying over the expression for
the heavy--light meson masses in the heavy-quark limit in QCD to our
would-be composite quarks, gives $M_{u_i^\prime}={\mathcal O}(
\Lambda_\text{HC}) + m_{\q_i} + m_{\mathcal S}$. Here
$\Lambda_\text{HC}$ is the HC confinement scale. The scalar ${\mathcal
  S}$ corresponds to a QCD heavy quark with mass between that of the
charm and the bottom quark. We thus estimate the HC-dynamics
contribution to the $u'_i$ mass to be roughly the $\rho$ mass,
similarly to what is found for the $D$ and $B$ mesons in QCD. We set
the would-be HC quark masses to
\begin{equation}
\begin{split}
   M_{u^\prime} &= M_{c^\prime} = M_\chi+
   m_{\q_1} +  m_{\mathcal S} \,,\\
   M_{t^\prime} &= M_\chi+ m_{\q_3} +  m_{\mathcal S} \,.
\end{split}
\label{eq:primemasses}
\end{equation}
For our benchmark this gives $M_{u',c'}=691$\,GeV and
$M_{t'}=718$\,GeV.

\section{Decay constants and couplings\label{app:AMD}}%

In this appendix we explain in more detail how we obtain our estimates
for the couplings of the HC resonances to the would-be composite
quarks, and our estimates for the decay constants of the would-be
composite quarks.

We start with the VMD estimates of the HC resonance couplings to the
would-be composite quarks, Eq.~\eqref{eq:rhouprime}.  The VMD
assumption states that the matrix element $\langle u' | J^{a\mu} | u'
\rangle$, where $J^{a\mu} = \bar \q_i \gamma^\mu ({T}^a)_{ij} \q_j\,$,
is dominated by the lowest-lying vector resonance $V^a$ with the same
quantum numbers as the current $J^{a\mu}$. For appropriately chosen
linear combinations of the flavor-group generators, $T^a$, these will
be the vector resonances $\rho, \omega, K^*, \phi$.  In the VMD limit
we can write
\begin{equation}
\langle u' | J^{a\mu} | u' \rangle \to \langle u' | u' V^a \rangle
\frac{1}{q^2 - M_{V}^2} \langle V^a | J^{a\mu} | 0 \rangle\, .
\label{eq:polology}
\end{equation}
Using the definitions in Eqs.~\eqref{rho:decay} and
\eqref{eq:rhouprime} we have
\begin{equation}
\langle u' | J^{a\mu} | u' \rangle \to g_{V}
\frac{f_{V}  M_{V}}{q^2 - M_{V}^2} \bar{u}^\prime {
  T}^a \gamma_\mu u^\prime, \, 
\end{equation}
where we have used $\sum \epsilon_\mu \epsilon_\nu^* = g_{\mu\nu} +
q_\mu q_\nu / M_{V}^2$ and the Dirac equation. On the other hand, the
vector current matrix element can also be written in terms of the form
factors
\begin{equation}\label{eq:ffgen}
\langle u' | J^{a\mu} | u' \rangle = \bar{u}^\prime {
  T}^a [\gamma_\mu f_1(q^2) + \frac{i\sigma_{\mu\nu}}{2m_{u'}} q^\nu
  f_2(q^2) ]u^\prime \, ,
\end{equation}
where the normalization condition is $f_1(0) = 1$. Equating the last
two expressions for $q^2 \to 0$ leads to the VMD relation
\begin{equation}
g_{V} =\frac{M_{V}}{f_{V}}\, .
\end{equation}

\begin{figure}[]
\begin{center}
\includegraphics{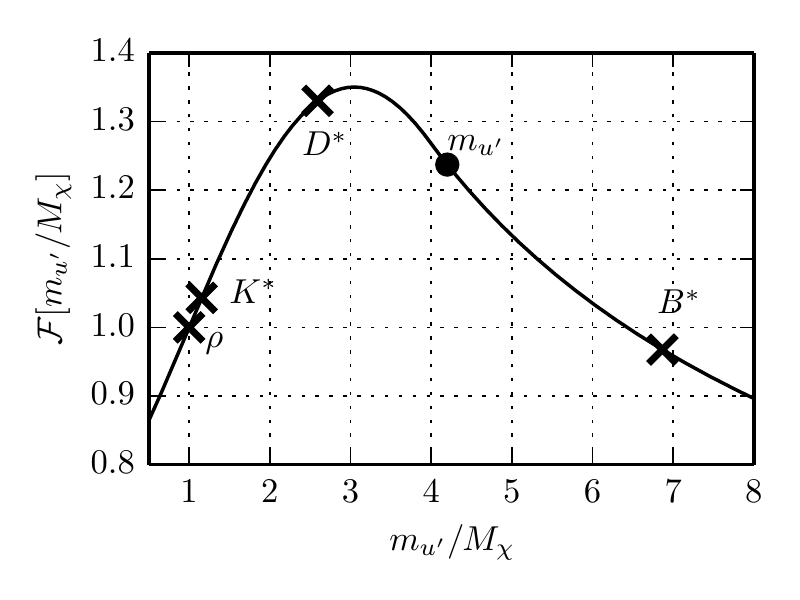}
\end{center}
\caption{The normalized interpolating function ${\mathcal F}$ used to
  estimate the decay constant $f_{u'}$ as a function of the
  dimensionless quantity $m_{u'}/M_\chi$. The function ${\mathcal F}$
  is chosen such that it reproduces the measured QCD decay constants
  $f_\rho$, $f_{K^*}$, $f_{D^*}$, $f_{B^*}$, denoted with crosses. The
  value of $m_{u'}$ used in the benchmark is denoted by a circle.
\label{fig:fit_f_n}}
\end{figure}

Next, we describe the determination of the decay constant, $f_{u'}$,
of the would-be composite quarks.  In this estimate we can take the
$\bar q_i$ to be massless, and focus entirely on the $m_{\cal S}$
dependence of $f_{u'}$. Again we are guided by QCD. For $m_{\cal S}\ll
M_{\chi}$ (i.e. $m_q\ll m_\rho$ in QCD) we use the fact that ChPT and
lattice QCD simulations show a linear dependence: $f_{u'}=a
(f_{u'})_0+b\,m_{\cal S} $. For heavy $m_{\cal S}\gg M_\chi$, HQET
yields the scaling $f_{u'}=c/\sqrt{m_{\cal S}}$.  Rescaling from QCD
yields an expression for $f^\text{HC}_{u'}$ of the form
\begin{equation} f^\text{HC}_{u'} = f_\rho^\text{QCD}
\frac{M_\chi}{M_\rho^\text{QCD}} {\mathcal F} \left(
\frac{m_{u^\prime}}{M_\chi}\right),
\label{eq:fu'} 
\end{equation} 
where we use the fact that the would-be composite quark $u'$ most
closely resembles the $\rho$ in QCD (i.e. it is a low-lying resonance,
but not a pNGB).  The dimensionless function ${\mathcal F}(x)$
captures the ChPT and HQET behaviours. For $x\to 0$ we have ${\mathcal
  F}(x)=c{\mathcal F}(0)+{\mathcal F}(0)' x$, while for $x\gg1$ we
have ${\mathcal F}(x)={\mathcal F}_\infty/\sqrt{x}$. The constant
${\mathcal F}_\infty$ is determined from the value of $f_{B^*}$ in
QCD. The HQET form of ${\mathcal F}(x)$ is assumed to be valid above
$x_{\rm match}=m_{u'}/M_\chi\simeq 4.1$, and a quadratic interpolation
is used for lower values of $x$, such that the QCD values of
$f_{D^*}$, $f_{K^*}$ and $f_\rho$ are all properly described. The
result is shown in Figure~\ref{fig:fit_f_n}.  For our estimates of
$f_{u'}$, only the HQET scaling turns out to be needed (along with the
proper rescaling to the HC scale), given the range of masses we
considered for $m_{\cal S}$.

\section{Decay widths\label{app:decay:widths}}%
The decay widths for $\rho$ and $K^*$ decays to pseudoscalar pairs are
given by
\begin{equation}\label{vector:decay:widths}
\begin{split}
{\Gamma_\rho \over m_\rho } &=  {1 \over  96 \pi} \bigg[  2 g_{\rho
    \pi\pi}^2 \left(1- {4 m_\pi^2 \over  m_\rho^2}  \right)^{3/2} \\ &
  \qquad \qquad + g^2_{\rho KK} \left(1- {4 m_K^2 \over  m_\rho^2}  \right)^{3/2} \bigg], \\
{\Gamma_{K^*} \over m_{K^*}  } &= {g_{K^* K \pi}^2 \over  64  \pi}  
\left[f_{\rm PS}\left(\tfrac{m_\pi}{ m_{K^*}}, \tfrac{m_K}{m_{K^*}} \right)\right]^3 \\
&\quad +\, {g_{K^* K \eta_8 }^2 \over  64  \pi}  
\left[f_{\rm PS} \left(\tfrac{m_{\eta_8}}{m_{K^*}},\tfrac{m_K}{ m_{K^*}} \right)\right]^3. 
\end{split}
\end{equation}
Here the phase-space function is defined as
\begin{equation}
f_{\rm PS}(x, y)=1-2 x^2- 2y^2\,.
\end{equation}
Note that, due to flavor-symmetry breaking, the effective couplings
$g_{\rho KK}$, $g_{K^* K \pi}$, $g_{K^* K \eta_8 }$ could differ from
$g_{\rho \pi\pi}$, defined in the flavor-symmetric limit in
Eq.~\eqref{Eq:WMD:grhopipi}. In our numerics we take, however, all of
them equal to the VMD value $g_\rho$, as discussed in
Section~\ref{subsec:vec:axial}.

The decay widths of $V_L$ and $V_H$ read:
\begin{equation}
\begin{split}
  {\Gamma_{V_L}\over m_{V_L}}= {g_{\rho}^2 \over 32 \pi }
  \sin^2\theta_V \left(1-4 {m_K^2\over m_{V_L}^2} \right)^{3/2}&\,,\\
  {\Gamma_{V_H}\over m_{V_H}}= {g_{\rho}^2 \over 32 \pi }
  \cos^2\theta_V \left(1-4 {m_K^2\over m_{V_H}^2} \right)^{3/2}&\,,
\end{split}
  \label{eq:mixed:dec:width}
\end{equation}
where $\theta_V$ is the mixing angle in Eq.~\eqref{VLH}, and we have
used the VMD estimate of the effective coupling.

The partial widths of the vector mesons to SM quarks are governed by
their mixing with the would-be composite quarks, see
Eq.~\eqref{eq:rhoqq}.  For instance, the $\bar K^{*1} \to \bar t u$
partial decay width is given by
\begin{align}
 \frac{\Gamma_{\bar K^{*1}\to \bar t u}}{M_{K^*}} & =
\frac{1}{16\pi}
f_{\rm PS}\bigg(0,\frac{m_t}{m_{K^*}}\bigg)
\Big[\big(\lambda_{13}^{R}\big)^2 +
\big(\lambda_{13}^{L}\big)^2 \Big] \nonumber\\
& \quad \times \left[ 1 - \frac{m_t^2}{2 m_{K^*}^2} 
  - \frac{1}{2} \bigg( \frac{m_t^2}{m_{K^*}^2} \bigg)^2 \right] \, , 
\end{align}
and similarly for $\Gamma_{\bar K^{*2}\to \bar t c} $ with
$\lambda_{13}^{L/R} \rightarrow \lambda_{23}^{L/R}$.

The ${}^3 P_1$ axial vectors can have $A \to VP $ decays and, if
kinematically allowed, $A \to VV$ decays.  The former usually
dominate.  To estimate the corresponding decay widths we follow the
phenomenological analysis for $A \to VP $ decays in QCD given in
Ref.~\cite{Roca:2003uk}, and carried out in the ${\rm SU}(3)$ limit.
The authors point out that in general one can express the decay
amplitudes in terms of two independent operators,
\begin{equation} 
\langle A_{\mu\nu}[V^{\mu\nu},P]\rangle \,,~~~\langle
A_\mu[V^\mu,P] \rangle\,.
\end{equation} 
They only choose the first operator for their study, which involves
the tensor representations for the vector operators.  However, they
explain that an analysis involving only the second operator would
yield similar results. The $A\to VP$ Lagrangian, in the ${\rm SU}(3)$
limit, is then
\begin{equation} {\cal
  L}_{AVP}=i\tilde{F} \langle A_{\mu\nu}[V^{\mu\nu},P]\rangle \,,
  \end{equation}
where $\langle\dots \rangle$ denotes a trace over the ${\rm SU}(3)$
generators, $\tilde F$ is a dimensionful coupling, and the factor $i$
insures that the Lagrangian is Hermitian.  The tensor-field operators
$V_{\mu\nu}$ is normalized such that
\begin{equation}
\langle 0|V_{\mu\nu}|V(P,\epsilon)\rangle =
\frac{i}{M_V}\left( P_\mu\,\epsilon_\nu-P_\nu\, 
\epsilon_\mu\right),
\label{eq:tensnorm}
\end{equation}
and similarly for $A_{\mu \nu}$.

The partial decay widths of the axial vectors are then given by
\begin{equation}
\Gamma_{A\to VP}=  \frac{|\lambda_{AVP}|^2 }{2\pi }\frac{\tilde{F}^2 q}{m_A^2}\left(1+
\frac{2}{3}\frac{q^2}{m_V^2}\right),
\label{eq:GAVP}
\end{equation}
where $q= f_{\rm PS}(m_V^2/m_A^2,m_P^2/m_A^2)\cdot1/{(2m_A)}$, and
$\lambda_{AVP}$ are decay-mode-specific dimensionless prefactors. For
instance, the mixed flavor octet-singlet states $A_L$ and $A_H$ decay
to $K^* K$ final states.  Summing over the $K^{* \pm} K^{\mp}$,
$K^{*0} \bar K^0 $, and $\bar K^{*0} K^0$ decay modes, the $A_L \to
K^* K$ and $A_H \to K^* K$ decay widths are obtained by using the
expression
\begin{equation}
\begin{split}
\lambda_{A_{L,H} K^* K}  =& 2 \Big({ \cos^2 \theta_{\rm A}^{\rm id} \over 2} + \sin^2 \theta_{\rm A}^{\rm id} \pm \sqrt{2} 
\sin \theta_{\rm A}^{\rm id} \cos \theta_{\rm A}^{\rm id} \Big)^{1/2},\\
\end{split}
\end{equation}
for the $\lambda$ prefactor in Eq.~\eqref{eq:GAVP}

The $a_1$ decays to $\rho \pi$ and $K^* K$. Summing over these final
states, the $a_1 \to \rho \pi$ and $a_1 \to K^* K$ partial decay
widths are obtained by using
\begin{equation} 
\lambda_{a_1 \rho \pi } =2\quad\text{and}\quad\lambda_{a_1 K* K } = \sqrt{2}
\end{equation}
in Eq.~\eqref{eq:GAVP}. The $K_1 \to \rho K, K^* \pi ,K \omega, K \phi
$ partial decay widths follow from Eq.~\eqref{eq:GAVP} using
\begin{equation} 
\begin{split}
\lambda_{K_1  \rho K ,K_1  \pi K^*} & = \sqrt{3\over
2}\,,\quad
\lambda_{K_1  \omega K  } = {1\over \sqrt 2} \,,\quad
\lambda_{K_1  \phi K  }  = 1\,.
\end{split}
\end{equation}

The fits to the QCD axial-vector decay widths yield a large allowed
range for $\tilde F$.  In particular, the solutions for $\tilde F$
vary from $\tilde F_{\rm QCD}\approx 1200$\,MeV to $\tilde F_{\rm QCD}
\approx 1600$\,MeV.  We rescale from QCD with the usual scale factor,
\begin{equation} \label{eq:app:FHC}
\tilde F_{\rm HC} = {M_{\chi} \over
    m_{\rho^{\rm QCD} } }\, \tilde F_{\rm QCD} \,.
\end{equation}

Finally, we quote the total decay width of the heavy scalar ${\cal
  S}$, including the phase-space factors
\begin{equation}
\begin{split}
\frac{\Gamma_{{\cal S} \to u_i \bar\q_i}}{m_{\cal S}} = 
\frac{|h_i|^2}{16\pi}
f_{\rm PS}\bigg( \frac{m_{u_i}}{m_{\cal S}}\,,
\frac{m_{\q_i}}{m_{\cal S}}\bigg)
                           \biggl(1 - \frac{m^2_{u_i}}{m_{\cal S}^2} -
                           \frac{m_{\q_i}^2}{m_{\cal S}^2}\biggl)\,.
\end{split}
\label{eq:app:Gammau':fulleq}
\end{equation}


\bibliographystyle{apsrev}
\bibliography{strong_int}

\end{document}